\documentclass[aoas]{imsart}

\RequirePackage{amsthm,amsmath,amsfonts,amssymb}
\RequirePackage{natbib}
\RequirePackage[colorlinks,citecolor=blue,urlcolor=blue]{hyperref}

\usepackage{enumerate}
\usepackage[utf8]{inputenc}
\usepackage{import}
\usepackage{multicol}
\usepackage{enumitem}
\usepackage{amsmath, amscd, amssymb, amsthm, xspace}
\usepackage{tabularx}
\usepackage{mathrsfs}
\usepackage[all]{xy}
\usepackage{subcaption}
\usepackage{float}
\usepackage{mathtools} 
\usepackage[toc,page]{appendix}
\usepackage{listings}
\usepackage{comment}
\usepackage{color}
\usepackage{makecell}
\usepackage{multirow}
\usepackage{caption}
\usepackage{subcaption}

\newtheorem*{theorem*}{Theorem}

\newtheorem*{prop*}{Proposition}

\newtheorem*{corollary*}{Corollary}

\newtheorem*{lemma*}{Lemma}

\newtheorem*{claim*}{Claim}
\newcolumntype{C}[1]{>{\centering\arraybackslash}p{#1}}
\newcolumntype{Y}{>{\centering\arraybackslash}X}

\newcommand{\Expect}[1]{\mathbb{E}\left[ #1 \right]}

\newcommand{\diffd}{\textnormal{d}}
\newcommand{\deriv}[2]{\frac{\diffd #1}{\diffd #2}}

\newcommand{\fourier}[1]{\mathcal{F}\left[#1\right]}

\newcommand{\norm}[1]{\left|\left|  #1 \right| \right|}

\newcommand{\mat}[1]{\mathbf{#1}}
\newcommand{\m}[1]{\mathbf{#1}}

\newcommand{\tr}[1]{\mathrm{tr}\left(#1\right)}

\newcommand{\degrees}{^{\text{o}}}

\newcommand{\rlang}{\texttt{R} }

\newcommand{\fraC}[2]{\frac{#1}{#2}}

\newcommand{\sneed}{\psi}

\begin{document}
\begin{frontmatter}

\title{A Physically-Motivated Stochastic Model of Tropical Cyclone Dynamic Wind Fields }
\runtitle{Physically Motivated TC Model}

\begin{aug}
\author[A]{\fnms{Matthew}~\snm{LeDuc}\ead[label=e1]{matthew.leduc@colorado.edu}\orcid{0009-0006-5892-8823}}
\author[A]{\fnms{William}~\snm{Kleiber}\ead[label=e2]{william.kleiber@colorado.edu}}
\address[A]{Department of Applied Mathematics, University of Colorado Boulder\printead[presep={ ,\ }]{e1,e2}}

\end{aug}

\begin{abstract}
Tropical cyclones (TCs) pose significant risk to infrastructure and human safety, and the frequency and intensity of large TCs are expected to increase in the future. 
Quantifying such risk to coastal and inland regions is challenging, largely due to the rare nature of TCs which limits historical data availability. 
We introduce a new stochastic model of TC wind fields conditional on a small number of physical variables commonly found in synthetic and observed storm track databases. 
The model consists of a physically-motivated mean trend and an expansion of the residual field into a wavelet frame. 
We demonstrate the approach on a dataset consisting of historical data for storms impacting the northwest Gulf of Mexico.
The model adequately captures asymmetry and nonstationary space-time correlation structures of observed wind fields.
\end{abstract}


\end{frontmatter}




\maketitle

\section{Introduction}\label{sec-intro}

Global climate change is expected to lead to an increase in the frequency and severity of severe weather events, particularly large tropical cyclones (TCs). 
The combined effects from storm surge, precipitation, and high winds are expected to lead to significant damages in coastal areas, both economically and in human mortality \citep{emanuel2005increasing,Dullaart2021,rosowsky2021projecting, ipcc2023}. 
Quantifying the potential damage due to TCs is a challenge, however, partly because historical observations are relatively scarce, due to the rare and extreme nature of TCs. 
As a result, risk assessment and infrastructure design increasingly rely on large ensembles of synthetic storms to characterize the range of plausible hazards \citep{iman2005projections,simiu2019wind}. 
While these ensembles can be generated using numerical weather prediction models such as the Weather Research and Forecasting model (WRF, \cite{wrfmodel}), the computational cost limits ensemble size and makes generating the samples necessary for probabilistic risk assessment impractical.

To meet this need, there has been considerable effort devoted to developing models of TC tracks and intensity (e.g. \cite{bloemendaal2020STORM,bosepintarLSTMhurricanes2022,nederhoff_tcwise,vickerytrackmodel}), as well as modeling and analysis of TC wind speed, direction, and geometry (e.g. \cite{abe2017tractable,foley2008statistical,Holland:1980,johnson2007fitting,murphy2025joint,vickery1995wind}, among many others). 
In addition, databases such as IBTrACS (\cite{ibtracs}) and HURDAT2 (\cite{hurdat}) contain data for historical storms. 
However, the data available in these track databases is limited and varies both between and within individual databases. 
Recently, machine learning approaches have been employed to estimate high-resolution wind fields based on environmental data \citep{yang2022machine}, however they often rely on environmental predictors such as wind shear, sea surface temperature, and humidity. 
While such geophysical variables are critical to the development and formation of TCs (e.g. \cite{gao2025crucial,gray1979hurricanes}), they are not contained in track databases, limiting the applicability of these methods for long term risk assessment. 
Other approaches, such as \cite{reich2007multivariatewinds}, are able to provide high resolution realizations of hurricane wind fields, however they require satellite observations and do not model the dynamic temporal evolution of the storm.

These limitations motivate a hybrid approach to the problem incorporating physical structure with statistical learning techniques \citep{kleiber2023stochastic,karniadakis2021physics,willard2022integrating}.
For geophysical systems, where observations and simulations are often limited relative to the dimensionality of the underlying processes, these approaches are particularly attractive. 
Rather than attempting to learn the full dynamics of a system directly from data, these methods exploit physical understanding to constrain the the system while allowing data-driven components to characterize unresolved processes and interactions. 
Our proposal falls in this framework, focusing on the development of a physically-motivated TC wind field model capable of providing stochastic simulations of the dominant structure at a manageable computational load while only relying on a minimal set of predictors derived from track data. 
Tropical cyclone wind fields contain strong physical structure arising from vortex dynamics, storm translation, environmental interactions, and surface effects, making them a natural candidate for this framework. 

We present a Lagrangian model (with respect to the storm center) for wind vectors using historical data generated by the Weather Research and Forecasting Model (WRF), and predictors derived from storm characteristics commonly available in databases such as IBTrACS and HURDAT2. 
We propose representing the storm-relative wind vector field in polar coordinates 
\begin{equation}\label{eq:windfield_model}
    \vec{w}(r,\theta,t) = \vec{\mu}(r,\theta,t)\exp\left( \sum_{j,k}{\beta}_{j,k}(t) \sneed_{j,k}(r,\theta) + {\varepsilon}(r,\theta,t)\right)
\end{equation}
where $r$ and $\theta$ are the radius and angle relative to the storm motion and $\vec{\mu}(r,\theta,t)$ represents a physics-guided model for the mean circulation that incorporates land interaction. 
The multiplicative term represents stochastic deviations from the mean circulation through a multiscale wavelet expansion, with $\varepsilon$ accounting for further unresolved variability. 
The resulting model can be viewed as a physics-guided stochastic emulator, where physically-interpretable components capture the dominant circulation, and statistical components describe unresolved processes and environmental interactions.


\section{Data}\label{sec:data}

 Our data are reanalysis fields of pressure and meridional and zonal winds from WRF. These are the same WRF runs as used for \cite{kleiber2023stochastic}, and the WRF parameters are described therein. The spatial grid has resolution 20 km $\times$ 20 km, and wind fields are reported every hour. Due to the relatively low resolution of the model, the wind field is relatively weaker and broader than the corresponding historical storm. First, for each time step in each storm we extract the wind speed at 10 m height and calculate the curl of the wind field from the meridional and zonal winds. The location of maximum curl is taken to be the center of the storm, and since these locations are restricted to the WRF grid a smoothing spline is applied to estimate the latitudes and longitudes as a function of time. From this time series we can calculate the translation speed and storm heading as a function of time to convert the Earth-relative meridional and zonal winds into storm-relative winds in the cross-track and along-track directions.  

From here, we are able to calculate the maximal wind speed and radius of maximal winds. We determine the maximal wind speed $V_m$ and the radius of maximum winds $R_m$, which is defined as the distance between the estimated eye and the grid cell corresponding to $V_m$. Any time with $V_m<10$ m/s or $R_m>350$ km is discarded. From this, the central pressure is calculated as the minimal WRF pressure within $3R_m/4$ km of the storm center and the time step is discarded if this value is greater than 1010 hPa. Given these values, we can calculate signed distance from the eye to the coastline, with negative distances denoting locations over land. Any time where the distance from the eye to the coastline is less than $-0.1 R_m$ is discarded.

At this point, we perform the final checks for usability. First, we check how much of the circle of radius 700km centered on the eye is within the WRF simulation domain and remove time steps where this quantity is less than $65\%$ or where $65\%$ or more of the same region is over land -- once TCs pass over land, the ingestion of dry continental air dramatically reduces storm intensity. 
The wind speeds over the ocean between $0.85 R_m$ and $1.15R_m$ are used to fit the model of angular variation, which is described in Section \ref{sec:mean_model_tcs}. If fewer than 20 datapoints satisfy this criterion or the available data cover a sector of width less than 120 degrees, the time step is discarded. We keep storms that have a record of 15 hours of valid time steps with no gap larger than 2 hours. The data in these gaps are imputed with a Kalman filter. In the case of Hurricane Harvey (2017), two time periods separated by about 24 hours satisfied these criteria, so  both are included in the training data individually. After this process, we are left with 19 storms described in Table \ref{tab:storminfo}. The storms cover a range of years from 1979-2017, and durations from 18-51 hours. 

\begin{table}[h!]
    \centering
    \begin{tabularx}{\textwidth}{|l|c|Y|Y|Y|}
    \hline
         Storm (Year) &Duration (Hours)&Max WRF wind (storm-relative, m/s)&Max WRF central pressure deficit (hPa)& Mean WRF Rmax (km) \\
         \hline
     Claudette (1979)&21&18.8&8.5&220\\
         \hline
         Elena (1979)&44&14.2&5.9&204\\
         \hline
         Allen (1980)&27&32.5&31.9&141\\
          \hline
        Alicia (1983)&29&20.0&6.8&169\\
        \hline
          Juan (1985)&39&22.7&30.6&216\\
        \hline
       Chantal (1989)&22&22.3&6.3&222\\
       \hline
         Jerry (1989)&20&17.5&5.6&248\\
         \hline
          Dean (1995)&19&17.9&7.6&274\\
          \hline
       Frances (1998)&51&21.7&15.6&251\\
       \hline
       Allison (2001)&23&20.2&6.3&234\\ 
       \hline
           Fay (2002)&45&19.0&9.2&178\\  
     \hline
     Claudette (2003)&33&22.3&6.6&195\\ 
     \hline
          Erin (2007)&18&17.9&5.8&212\\  
          \hline
       Edouard (2008)&19&20.6&3.9&141\\  
       \hline
           Ike (2008)&20&38.4&30.2&145\\ 
           \hline
          Bill (2015)&30&21.4&5.4&184\\
          \hline
         Cindy (2017)&22&20.9&14.7&201\\
         \hline
      Harvey 1 (2017)&20&26.8&17.3&164\\
      \hline
      Harvey 2 (2017)&45&18.2&13.7&125\\  
         \hline
    \end{tabularx}
    \caption{Storm names along with duration and environmental parameters derived from WRF. The central pressure deficit is defined as the difference between 1013 hPa and the central pressure from WRF.}
    \label{tab:storminfo}
\end{table}

\section{Methods}\label{sec:methods}
In this section, we present the statistical model to be used in the emulator. We will first discuss the bulk flow model, which governs the large-scale circulation and the interaction of the storm with land, in Section \ref{sec:mean_model_tcs}. The model for the small-scale structure, which is based on a multiresolution expansion of the data after the bulk flow model is removed, is introduced in Section \ref{sec:resid_model}.

\subsection{The Bulk Flow Model}\label{sec:mean_model_tcs}

\begin{figure}[h!]
    \centering
    \includegraphics[width=0.65\linewidth]{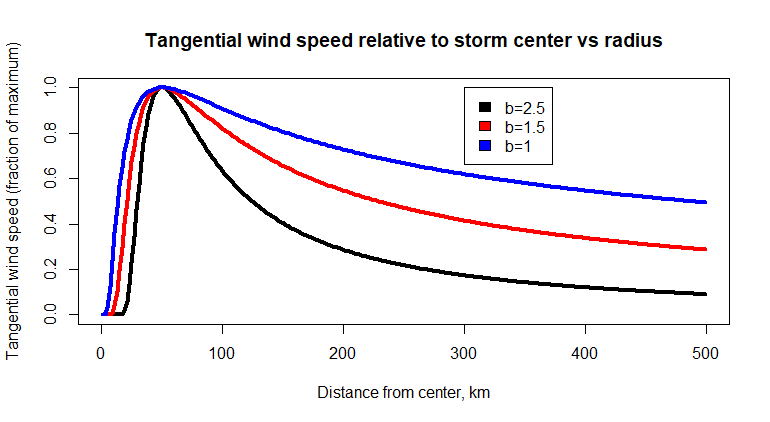}
    \caption{Tangential winds vs radius for the Holland wind-pressure model for varying $b$, assuming $x(r)=1/2$.}
    \label{fig:holland_tangential_winds}
\end{figure}

The model of the bulk flow is based on the Holland model, a classical vortex model often used in studies of tropical cyclones (\cite{Holland:1980,holland2010revisedwinds}). The model proposes that the tangential wind speed as a function of the distance from the eye satisfies a gradient balance and parameterizes the pressure deficit as a function of distance $r$ from the eye as
\begin{equation}
    \label{eq:cpd_holland}
    \delta_p(r) = (p_a-p_c)\exp\left(-\left(\frac{R_{m}}{r}\right)^b\right)
\end{equation}
where $p_c$ is the pressure at the center, $p_a$ is the ambient pressure (assumed to be 1013 hPa), $R_{m}$ is the radius of maximal winds in km, and $b$ is a shape parameter. The Holland model yields the tangential wind profile (\cite{holland2010revisedwinds,hu2012consistencywindforecasting})
\begin{equation}\label{eq:hollandmodel}
\begin{split}
     V(r) = &\left(\frac{100b}{\rho e^1} \left(\frac{R_m}{r}\right)^b \delta_p(r)+\left(\frac{rf}{2}\right)^{2} \right)^{1/2}-\left(\frac{rf}{2}\right) \\
            b= &\frac{\rho e^1\left(V_{m}^2+V_{m}R_{m}f\right)}{100\Delta p}
            \end{split}
\end{equation}
where $\Delta p=p_a-p_c$ is the central pressure deficit in hPa, $\rho\approx 1.225kg/m^3$ is the density of air, $f$ is the Coriolis parameter, and $V_m$ is the maximum wind speed (in m/s). 
The expression for $b$ can be derived by noting that when $r=R_m$, $V(r)=V_m$, and expressions for approximating these quantities based on other data exist (e.g. \cite{chavas2025simple,holland2008pressurewind, nederhoff2019estimates}).  An example of these tangential winds for varying $b$ parameters is shown in Figure \ref{fig:holland_tangential_winds}. For the purposes of this model, we will assume that these data are all available. These are the typical physical parameters used to describe a tropical cyclone, and any standard track database such as IBTrACS will include $\Delta p, V_m$, and $R_m$ at the minimum.

\subsubsection{Modeling the Angular Portion}

While the Holland model is commonly used and physically well-motivated, it is a model for a scalar quantity and also enforces radial symmetry in the wind field. However, tropical cyclone wind fields are vector fields which are highly variable in angle, which motivates more structured models. In order to introduce this additional structure, we decompose the mean wind field as
 \begin{equation}\label{eq:mn_vector_field}
     \vec{\mu}(r,\theta,t) = \tilde{V}(r,t)\vec{T}(\theta, t)
 \end{equation}
 where $\tilde{V}(r,t)$ is the generalized Holland model  normalized to unit maximum wind speed. We model the angular portion $\vec{T}(\theta, t)$ as
\begin{equation}
    \label{eq:T_vector}
    \vec{T}(\theta, t) = \begin{pmatrix}
        T_{\perp}(\theta, t) \\T_{\parallel}(\theta,t)\end{pmatrix}=\sum_{k=0}^N\left[\vec{a}_{k,1}(t)\cos(k\theta)+\vec{a}_{k,2}(t)\sin(k\theta)\right]
\end{equation}
where $T_{\perp}$ and $T_{\parallel}$ are winds in the cross-track and along-track directions and the maximal wavenumber $N$ is a model parameter. We set $N=2$, which captures the dominant shape of the asymmetry while avoiding overfitting of small oscillations. A similar model was applied in \cite{ModelingandParameterEstimationofHurricaneWindFieldswithAsymmetry} to model the scalar winds using information about the storm translation and wind shear, however this model only considered the total wind speed rather than the wind vectors. Additionally, shear data are not typically found in tropical cyclone track databases, complicating estimation. 

The structure of the angular portion of the model suggests the possibility of correlation between the Fourier vectors, both between entries of the vector and among the vectors themselves, which challenges many standard estimation routines for vector autoregressions that assume that the components are uncorrelated at time $t$ given the history of the series. One way to avoid this is to apply the ``curds and whey'' method 
(\cite{breiman1997predicting,kidd2014implementation}), which applies a shrinkage operator in canonical coordinates to improve estimation in the case where this assumption does not hold. 
However, the curds and whey method relies upon a linear regression, which limits the ability to learn the complex dependencies between the predictors and responses which are a feature of this dataset. For this reason, we prewhiten the sample coefficients by their empirical covariance matrix and adopt a generalized additive model (GAM) form for the coefficient vector, where for storm $i$ coefficient $j$ takes the form
\begin{equation}\label{eq:coeff_mn}
    \Expect{a_j(t)|D_i} = c_{i,j}+\sum_{x\in D_i} f_x(x(t))
\end{equation}
where $D_i$ is the set of predictors (including cross-terms) for storm $i$ and $c_{i,j}$ are stormwise random effects for each coefficient. We choose to model $f_x(x)$ as a thin plate regression spline, which are low-rank approximations to thin plate splines \cite{wood2003thin}. 

While this model is able to generate realistic estimates for the coefficients given the predictors, there is additional variability that is unaccounted for in this model. We can write the model for the coefficients as a vector valued stochastic process with expectation given by Eq. \eqref{eq:coeff_mn} as 
\begin{equation}\label{eq:avec_expectation}
    \vec{a}(t)|D_i = \Expect{\vec{a}(t)|D_i} + \vec{Z}(t)
\end{equation}
where $\vec{Z}(t)$ is a mean-zero stochastic process. Since the residuals after the estimation of the mean retain temporal correlation, we choose to model $\vec{Z}(t)$ as a hierarchical state-space process
\begin{equation}\label{eq:avec_resids}
    \begin{split}
        \vec{Z}(t) =& \mat{G}(t)\ast \vec{\eta}(t)\\
        \diffd \hat{\eta}(\omega)=& S(\omega)^{1/2}\diffd \hat{W}(\omega)\\
        W(t)\sim& \mathcal{MVN}(0,\mathbf{\Sigma})
    \end{split}
\end{equation}
where $\mat{G}$ is a causal impulse response function, $\vec{\eta}$ are innovations, and $\ast$ is a time domain convolution. This is a similar structure to a standard vector autoregression, with the exception that we allow the innovations $\vec{\eta}$ to be temporally correlated. This modification is motivated by the underlying physics of the problem. The unresolved processes that influence hurricane structure and motion, such as environmental steering flow, shear-induced asymmetries, and other storm-environment interactions, evolve on finite temporal scales and therefore are not well-represented by temporally independent perturbations. Allowing temporally correlated innovations provides a mechanism for representing unresolved forcing with finite temporal coherence, leading to simulated trajectories with more realistic smoothness and variability while preserving the causal dynamical structure encoded by the impulse response function.

\subsubsection{Decay Over Land}

A classic model for the decay of winds over land is the exponential decay model described in \cite{kaplan1995landwinddecay}. The proposed model consists of an exponential decay term and an instantaneous drop in winds upon hitting the coast. They provide empirical values for the Gulf coast region and model the decay of winds over land as
\begin{equation}\label{eq:wind_decay}
\begin{split}
    \deriv{V}{t} =& -\alpha F_{land} V,\ V(0) = RV_0\\
    \alpha=& 0.095,\ R=0.9
    \end{split}
\end{equation}
where $F_{land}$ is the fraction of the circulation from the Holland model over land and $t$ is the time since landfall. While simple, the model is local in that the effects of landfall are limited to the winds which are over land, while the effects of landfall are known to be nonlocal (e.g. \cite{SimpleModeloftheTropicalCycloneBoundaryLayeratLandfall,subramanian2016modeling}). For that reason, we seek to replace this with a nonlocal landfall operator that is capable of adjusting the shape and intensity of the circulation of the tropical cyclone in accordance with its interaction with land. Much like the model of \cite{kaplan1995landwinddecay}, we consider this to be a multiplicative effect on the entire wind field, and we write the mean wind field as
\begin{equation}
    \label{eq:mean_wind_with_decay}
    \vec{\mu}(r,\theta, t) = \tilde{V}(r,t)\vec{T}(\theta, t)\exp(\delta_\mu(r,\theta,t))
\end{equation}
with $\delta_\mu(r,\theta, t)$ parameterized by a feedforward neural network. The network is designed to learn a nonlocal landfall operator by mapping storm-scale and land interaction predictors to a multiplicative correction of the mean wind field. Inputs to the network include the storm predictors used in the model for the Fourier coefficients, together with quantities describing the interaction of the circulation with the coastline, such as the fraction of the circulation over land and additional geometric measures of land exposure. These predictors are chosen to allow the network to account for the nonlocal influence of landfall on the storm circulation. The details of the network and predictors will be discussed in Section \ref{sec:decay_land_fitting}.

\subsection{The Small-Scale Model}\label{sec:resid_model}

This section describes the model of small-scale variation present in the data after the removal of the bulk-flow model described in the previous section. First we will discuss the low-rank portion, which consists of a multiresolution expansion, and then how the field covariance is modeled. 

\subsubsection{The Solid Needlet Frame}
We base our model of the residuals on an expansion into the \textit{solid needlet} frame. The solid needlets are wavelets supported on the unit ball first described in \cite{petrushev2008localizedframesball}, and are an adaptation of the widely used class of wavelets known as needlets (\cite{needletintro}) to the unit ball. Needlets have been widely used, especially in astronomy and atmospheric science, due to their excellent spatial and harmonic localization (e.g. \cite{kerkyacharian2007needlet,cmbneeds,basak2012needlet,axingneed,yan2018estimating,tongyithesis}). Since needlets are compactly supported in harmonic space and decay faster than any rational function in physical space, they provide an easily interpretable partition of the modeled process into spatially and spectrally localized components. Additionally, they form a Parseval-tight frame, meaning that the coefficients satisfy Parseval's Identity (\cite{Intro_frames}). Solid needlets have these same properties, as shown in \cite{petrushev2008localizedframesball}. 
For fields on the unit ball in $\mathbb{R}^2$ they take the form
\begin{equation}\label{eq:d2_wavelets}
\begin{split}
\sneed_{-1,0}(x) &=\frac{1}{ \sqrt{2\pi}}\\
   \sneed_{j,k}(x) &= \frac{\sqrt{w_{j,k}}}{\pi}\sum_{n=0}^{\infty}b\left(\frac{n}{B^j}\right)\sum _{i=0}^n U_{n,i}(x)U_{n,i}(\vec{\xi}_{j,k}) \\
     U_{n,i}(r,\theta) &= \frac{\sin\left((n+1)\arccos\left(r\cos(\theta-\frac{i\pi}{n+1}\right)\right)}{\sin\left(\arccos\left(r\cos(\theta-\frac{i\pi}{n+1}\right)\right)}
    \end{split}
\end{equation}
which can be derived from the form in \cite{petrushev2008localizedframesball} using the results about the reproducing kernel for polynomials on the disk in \cite{xu1999summability} and \cite{waldron2008orthogonal}.
Because solid needlets are spatially localized, coefficient estimates are robust to discrepancies between the observed and true field that occur away from the center. Consequently, coefficients associated with well-observed portions of the storm can still be estimated reliably even when part of the storm lies outside the domain. This is shown quantitatively for the spherical needlets in Proposition 10.9 of \cite{RFSphere}, and an analogous result applies to the solid needlets as well (\cite{mythesis}).

For the purposes of this work, we choose to set the maximum value of $j=3$. The number of wavelets at each level of resolution corresponds to the size of the associated quadrature rule. We used the disk quadrature rules derived in \cite{takaki2022schemes} for construction of the basis. Their construction, based on selecting rules that best satisfy an efficiency criterion, leads to a set of rules with relatively few quadrature nodes. Using these rules leads us with less than 300 wavelet coefficients to model in each direction, a manageable task. Increasing $j$ further increases the size of the problem substantially (setting $j=4$ introduces an additional 732 wavelets), and when $j=3$ we are able to capture 80\% or more of the total variability in the residuals depending on the storm. For this reason, increasing $j$ further is not considered necessary.

\subsubsection{The Covariance Structure}

Lastly, we must model the leftover, small-scale variation $\varepsilon(r,\theta,t)$. There are several ways to do this, and the choice of model informs how we will fit the residual coefficients. We assume that the solid needlet expansion is able to model the large-scale correlation structure, and leave $\varepsilon(r,\theta,t)$ to describe local structure. This is exactly the model setup that motivates many spatial Gaussian process models, such as \cite{frk,sangfullscale,nychka,katzfuss2017,bgl,fsbgl}, and many others. We will, for the purposes of this work, assume that the covariance kernel of $\varepsilon(r,\theta,t)$ is either a delta function, in which case $\varepsilon(r,\theta,t)$ is a white noise, or compactly supported in space, allowing for significant computational speedup especially on datasets with large numbers of spatial locations. Combining this with the random coefficients in the solid needlet expansion leads to a flexible covariance model that naturally handles multiple scales of variability.

\section{Estimation}\label{sec:model_est}
\subsection{Estimation of the Bulk Flow Model}\label{sec:fit_fouriercoeffs}

Fitting the bulk flow model proceeds in two distinct steps. since the Holland model is defined directly by the parameters in the track database, we focus on fitting the process model for the coefficients $\vec{a}(t)$ given by Eqs. \eqref{eq:avec_expectation} and \eqref{eq:avec_resids}. First, we will discuss the predictors chosen, then the process for fitting the deterministic and stochastic portions of the model.

\subsubsection{Choosing Predictors}\label{sec:choosing_preds}
This project is motivated by a desire to fit a stochastic model of TC winds using a minimal number of physical variables that are commonly found in tropical cyclone databases. The minimal variables contained in those databases are the storm location, maximum wind speed, radius of maximal winds, and central pressure deficit at a series of time points. We also assume that the modeler has access to a database containing information about which locations correspond to land and open water. While this is a small number of variables, they can be used to get a significant amount of insight into the behavior of the storm. The predictors and their physical interpretation are summarized in Table \ref{tab:mean_model_preds}. Due to the small size of the simulation domain we do not include latitude and longitude of the eye as predictors, however a similar model meant for application over a larger domain might find that they are impactful. 

\begin{table}
\centering
\renewcommand{\arraystretch}{1.2}
\begin{tabular}{|c|c|c|}
\hline
\textbf{Storm Intensity} &
\textbf{Storm Geometry} &
\textbf{Storm Motion} \\
\hline
$V_m$, $\dot{v}_{max}$ &
$R_m$ &
$v_t$ \\
Max. wind speed &
Radius of max. wind &
Translation speed \\
\hline
$\Delta p$, $\dot{\Delta}p$ &
$d_c$ &
$\dot v_t$ \\
Pressure deficit &
Distance to coast /
$R_m$ &
Along-track acceleration \\
\hline
$F_{land}$&
$\left(g_{||},g_{\perp}\right)$
&
$v_t\dot{\phi}$ \\
Fraction over land
&Coastal interaction
&
Cross-track acceleration \\
\hline
\end{tabular}
\caption{Predictors used in the asymmetry model.}
    \label{tab:mean_model_preds}
\end{table}

The primary predictors are the storm translation speed $v_t$, which can be estimated from the time series of storm locations, the central pressure deficit $\Delta p$, and the maximum wind speed and radius of maximum winds $V_m$ and $R_m$, as well as the time derivatives of the first two. Using the coast database, we are also able to derive several other predictors. The first is the ratio of the signed distance from the eye to the coast with $R_m$, called $d_c$. A positive value means that the eye is over the ocean, and a negative value means that it is over land. The second, $F_{land}$, is given as the percentage of the wind circulation as defined by the Holland model which is over land. These predictors describe how much of the storm is over land as well as whether the storm is able to extract additional heat and moisture from the ocean. The last two land interaction predictors are given by $g_{||}$ and $g_{\perp}$, which describe the local interaction of the storm with the coastline. They are computed by averaging the gradient of a smoothed land indicator field over the storm's inner-core region and resolving the resulting vector into components parallel and perpendicular to the storm motion. These predictors distinguish between coastlines encountered head-on and coastlines encountered tangentially, allowing for variations in the orientation of the coast near landfall to affect the dynamics of the model \citep{hlyiwakasymmetrylandfall}.

To derive the last two predictors, we consider the storm to be a rigid body moving along the track with velocity vector 
\begin{equation}
    \vec{v}(t) = \begin{pmatrix}
        v_t\cos(\phi) \\v_t\sin(\phi)
    \end{pmatrix}
\end{equation}
where $\phi$ is the bearing relative to north. Then by the product rule we have
\begin{equation}
    \deriv{\vec{v}}{t} = \deriv{v_t}{t}\begin{pmatrix}
         \cos(\phi) \\\sin(\phi)
    \end{pmatrix}+v_t \deriv{\phi}{t}\begin{pmatrix}
         -\sin(\phi) \\\cos(\phi)
    \end{pmatrix}
\end{equation}
which gives the acceleration of the rigid body induced by the curvature in the path in terms of its component along the direction of the storm path ($\deriv{v_t}{t}$) and perpendicular to that direction ($v_t\deriv{\phi}{t}$). Letting $\kappa=\frac{1}{v_t}\deriv{\phi}{t}$ be the curvature of the storm track, we can see that this second term is $v_t^2\kappa$, the centripetal acceleration due to the curvature. Because of their roles in the physics of the bulk motion of the flow, there are our last two predictors. We additionally include a stormwise random intercept and tensor splines for interactions between $v_t$ and $V_m$ as well as between $V_m$ and $\Delta p$, which have a well-established physical relationship (e.g. \cite{chavas2025simple}).

\subsubsection{Fitting the Fourier Coefficients}

\begin{figure}[h!]
    \centering
    \begin{subfigure}[b]{0.85\textwidth}
        \centering
        \includegraphics[height=2.5in]{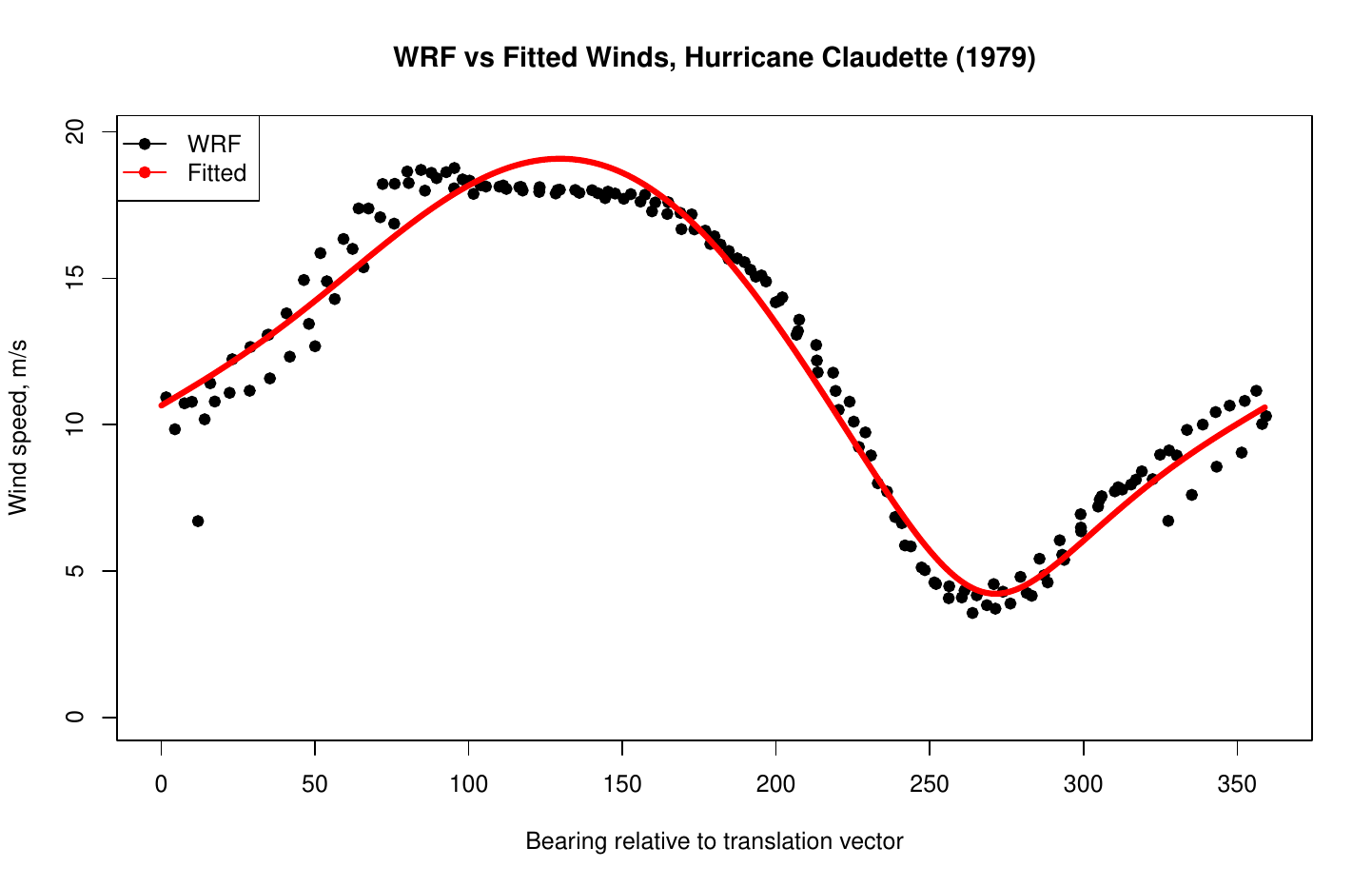}
    \end{subfigure}\hfill
    \begin{subfigure}[b]{0.85\textwidth}
        \centering
        \includegraphics[height=2.5in]{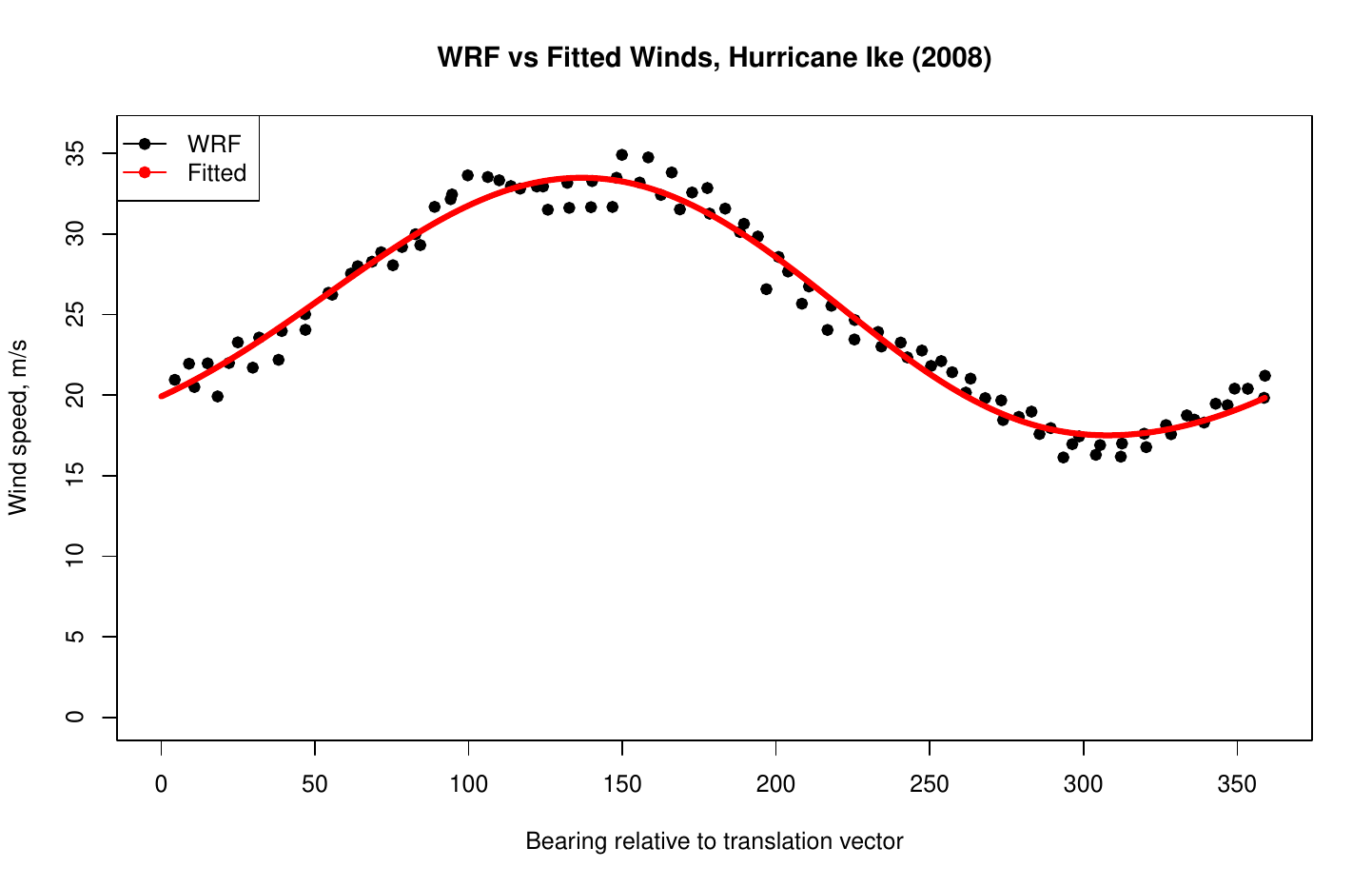}
    \end{subfigure}
    \caption{Two examples of winds from WRF compared to the model $\vec{T}(\theta, t)$ fit via nonlinear least squares. The fitted winds seem to match the data well.}\label{fig:wrf_v_T}
\end{figure}

Once the predictors are chosen, we proceed to fitting the Fourier coefficients. At each suitable timestep in each storm, defined as timesteps where more than 65\% of the storm is inside the simulation domain, we take the wind speed and bearing relative to the translation vector at every location over water within $R_m\pm15\%$ of the eye. The coefficient vector $\vec{a}(t)$ is then fit via nonlinear least squares as 
\begin{equation}
    \label{eq:fit_avec_to_wrf}
    \begin{split}
    \vec{a}(t)&=\mathrm{argmin}\sum_{i}\left(\|\vec{w}_{WRF}(\theta_i,t)\|-\|\vec{T}(\theta_i,t)\| \right)^2+\sum_k\lambda_k \left(\norm{\vec{a}_{k}}^2+\norm{\vec{b}_{k}}^2\right)
    \end{split}
\end{equation}
with $\vec{T}(\theta, t)$ given by Eq. \eqref{eq:T_vector}. This procedure is able to generate accurate fits to the observed angular profiles, see Figure \ref{fig:wrf_v_T} for examples from Hurricane Claudette (1979) and Hurricane Ike (2008). 

Once the coefficients have been calculated for all feasible time steps and storms, we proceed to fitting the mean model. The first step is to whiten the coefficient vectors via their empirical covariance to reduce cross-coefficient dependence in the mean structure, then proceed to fitting the model. The fitting is done using restricted maximum likelihood implemented via the \rlang package \texttt{mgcv} \citep{mgcv} assuming that the coefficient means each take the form of Eq. \eqref{eq:coeff_mn}, that is, a stormwise random intercept and thin plate regression splines for the remaining predictors. This nonparametric fitting method allows for a more flexible, data-driven modeling approach to the coefficient vectors at the cost of some computational complexity. This tradeoff, however, is worth it in order to be able to capture the complex, nonlinear dependencies between our predictors and the Fourier coefficients.

\subsubsection{Fitting the Stochastic Portion}
In the previous section, we discussed how to fit the expectation of the asymmetry coefficients using GAMs. In this section, we focus on the stochastic portion of that model, which is the process $\vec{Z}(t)$ in Eq. \eqref{eq:avec_resids}. We consider the equivalent model in the frequency domain given by
\begin{equation}
    \hat{Z}(\omega)=\hat{G}(\omega)\hat{\eta}(\omega)
\end{equation}
where $\hat{Z}(\omega)$ is the vector $(\hat{Z}_1(\omega),...,\hat{Z}_6(\omega))^T$ and $\hat{Z}_j(\omega)=\fourier{Z_j(t)}$ is the Fourier transform of $Z_j(t)$, with $\hat{\eta}$ defined similarly as the Fourier transforms of the innovations and $\hat{G}(\omega)$ is the transfer function given by
\begin{equation}
    \hat{G}(\omega)=\sum_{h}G_{h}
e^{-i\omega h}.
\end{equation}

Our task then is to estimate $\hat{G}$, $S_{\eta}(\omega)$ the power spectrum of the innovations, and $\mat{\Sigma}$ the cross-correlation of the innovations, jointly via Wiener filtering (\cite{wienerfilter}). To initialize the algorithm, we fit a first order vector autoregression to the residuals from each storm generated after fitting the GAMs and compute initial transfer function $\hat{G}^{(0)}(\omega)$ and cross-correlation matrix $\mat{\Sigma}^{(0)}$, and assume a flat power spectrum for $S_{\eta}^{(0)}(\omega)$ corresponding to temporally uncorrelated innovations.

At step $n$, the estimate of the covariance of the observations $\hat{Z}_i(\omega)$ for storm $i$ is given by
\begin{equation}
    \label{eq:obs_covariance}
    S_{Z,i}^{(n)}(\omega)
=
\hat{G}^{(n)}(\omega)
S_{\eta}^{(n)}(\omega)
\hat{G}^{(n)}(\omega)^H
+
S_\epsilon(\omega)
\end{equation}
where \(H\) denotes the Hermitian transpose, $S_{\eta}(\omega)=S_{\eta}(\omega)\mat{\Sigma}$, and $S_\epsilon(\omega)$ is an error covariance spectrum assumed to be flat and diagonal. The Wiener estimator of the innovations is then
\begin{equation}
    \label{eq:estimate_eta}
    \hat{\eta}_i^{(n+1)}(\omega)
=
S_{\eta}^{(n)}(\omega)
\hat{G}^{(n)}(\omega)^H
S_{Z,i}^{(n)}(\omega)^{-1}
\hat{Z}_i(\omega).
\end{equation}
The estimates $\hat{\eta}_i^{(n+1)}(\omega)$ are then used to calculate $\mat{\Sigma}^{(n+1)}$, which is then used to estimate $S_{\eta}^{(n+1)}(\omega)$. 

The final step is then to estimate $\hat{G}^{(n+1)}(\omega)$ as the impulse response that best fits
\begin{equation}
    \hat{Z}_i(\omega)=\hat{G}^{(n+1)}(\omega)\hat{\eta}_i^{(n+1)}(\omega).
\end{equation} 
over all the storms.
This is a multivariate deconvolution problem and thus can be ill-conditioned, especially at frequencies and lags where the signal to noise ratio is small. This can lead to unstable impulse response estimates with substantial oscillation between temporal lags. For this reason, we introduce two penalties to the model. The first is a ridge penalty, and the second is a Sobolev penalty to suppress rapidly varying lag-to-lag structure in the estimate. From a physical perspective, this reflects the assumption that the large-scale dynamical adjustment of the hurricane wind field to unresolved forcing evolves smoothly, implying that the associated impulse response should vary smoothly across lags. The Sobolev penalty takes the form of a penalty on either the time or frequency derivative of the impulse response, which can be formulated as a finite difference penalty on the impulse response if the chosen penalty is on the time derivative or a penalty on large lag terms if a penalty on the spectral derivative. The estimation can be performed by solving the augmented least squares problem
\begin{equation}
    \label{eq:irf_least_squares}
   \hat{ \mathbf{G} }= \text{arg}\underset{\mathbf{G}}{\text{min}}\norm{ \begin{bmatrix}
        \mathbf{Z}\\\mathbf{0}
    \end{bmatrix} - \begin{bmatrix}
        \mathbf{H}\\\mathbf{L}
    \end{bmatrix} \mathbf{G}}_F^2
\end{equation}
where the matrix $\mathbf{Z}$ contains the residuals from the coefficient estimation for all storms, $\mathbf{H}$ is a matrix of time domain innovations, and $\mathbf{L}$ is the Cholesky factor of the penalty matrix $\mathbf{L}^T\mathbf{L}=\lambda_r\mathbf{I}+\lambda_s\mathbf{P}_s$, with $\mathbf{P}_s$ the penalty matrix associated with the Sobolev penalty. This can be solved using QR decomposition, which gives a numerically stable solution without forming the normal equations. The result is, for $K$ Fourier coefficients and $H$ lags, a $KH\times K$ matrix of the form
\begin{equation}
    \hat{\mathbf{G}}=\begin{bmatrix}
        \hat{\mathbf{G}}_0\\ \hat{\mathbf{G}}_1\\\ \vdots \\\ \hat{\mathbf{G}}_H
    \end{bmatrix}
\end{equation}
where each block $ \hat{\mathbf{G}}_h$ describes the contribution of the innovations at lag $h$ to the evolution of the Fourier coefficients.

Once the impulse response is estimated, the algorithm is terminated when the impulse response and innovation cross-correlation stop changing. Then, a generalized Mat\'ern power spectrum of the form
\begin{equation}
    \hat{S}(\omega)=\frac{\sigma^2}{\left(\kappa^2+\omega^2\right)^{\nu}}
\end{equation}
is fit to the estimated power spectrum, providing a parsimonious model of variability in frequency and avoiding overfitting to the residual variation.

Because  the spectral density can be factored as seen in Eq. \eqref{eq:obs_covariance}, the spectrum of the residuals and the impulse response are not independently identifiable without additional restrictions. However, due to our restrictions on the form of the power spectrum, they play different roles in the temporal evolution of the model. While the spectrum describes common dynamics in the frequency domain, the impulse response function provides lag-dependent corrections to these dynamics, allowing the temporal dependence to vary across lags while retaining the common spectral structure. The spectrum therefore defines the baseline temporal dynamics, while the impulse response function modulates these dynamics through lag-dependent corrections.

\subsubsection{Decay Due to Land Interaction}\label{sec:decay_land_fitting}

We fit the multiplicative correction $\delta_\mu(r,\theta,t)$ to the mean wind field including the exponential decay from land interaction 
using a one-dimensional convolutional neural network with the $\texttt{torch}$ package in \rlang \citep{torchRpackage}. The network takes the available predictor channels as input and applies two hidden convolutional layers with 32 and 64 channels, respectively, using convolutional kernels of width three and ReLU activation, followed by a linear output layer with two channels, one for $\delta_\mu$ and one for the residual standard deviation $\sigma$. The convolutions are applied along the temporal dimension, with the same network shared across spatial locations. This allows the model to learn nonlinear relationships between the storm characteristics, land interaction, and the resulting deformation and attenuation of the wind field. 
Based on the predictors, we model the observed wind speed $w(r,\theta,t)$ as
\begin{equation}
    \ln(w(r,\theta,t))\sim \mathcal{N}(\ln(\mu(r,\theta,t))+\delta_\mu(r,\theta,t),\sigma(r,\theta,t) )
\end{equation}
with the likelihood conditionally independent across spatial locations. The network learns the expressions for $\delta_\mu$ and $\sigma$ jointly, allowing the model to represent both the effects of land interaction and the uncertainty in the representation. We constrain $\sigma$ to be positive by means of a softplus transformation. This formulation for the model maintains the physically motivated baseline $\mu(r,\theta,t)$ while still leveraging the flexibility of neural networks to learn departures from it. 

A quadratic penalty $0.001\,\delta_\mu^2$ is added to the negative log-likelihood to discourage unnecessarily large departures from the baseline mean field. Parameters are estimated using the Adam optimizer with a learning rate of $5\times10^{-4}$ for 100 epochs, with each storm considered its own mini-batch. The final-layer weights and biases are initialized to zero, giving an initial mean correction of zero and a common initial standard deviation of $\ln2$. 

The network relies on an expanded set of predictors relative to the ones provided in Table \ref{tab:mean_model_preds}. In addition to using the values of $g_{\parallel}$, $g_{\perp}$ and distance to the coast calculated at each spatial location, the network takes in the ratio between the distance from each spatial location to $R_m$, the values of $r\sin(k\theta)$ and $r\cos(k\theta)$ for k=1,2, the value of the mean wind field, a weighted average of $F_{land}$ since the start of the storm, and the smoothed land mask from which $g_{\parallel}$ and $g_{\perp}$ are calculated. These predictors allow the network to learn global corrections from the values of the physical variables as well as local corrections due to the values of the location-specific predictors to reduce bias from the mean model and estimate the uncertainty in the estimates.

\subsection{Fitting the Small-scale Process}

The model for small-scale variability is fit in three steps. The first step arises when the neural network is fit to learn the land decay. At this point, the model additionally learns the marginal standard deviations at each spatial location as a function of the predictors. 
Once these are learned and the mean model is fit, the log-residuals are standardized by the learned standard deviations and fit using several different spatial Gaussian process models assuming that the fields have no temporal correlation. 

Since the spatial correlation structure is expected to be nonstationary, we consider four different models for the spatial structure that naturally incorporate it: Fixed Rank Kriging (FRK ,\cite{frk}), LatticeKrig (LK, \cite{nychka}), the Basis Graphical Lasso (BGL, \cite{bgl}), and the Full Scale Basis Graphical Lasso (FSBGL, \cite{fsbgl}). 
These methods propose that the random field can be decomposed into the form 
\begin{equation}
    Y(s,t) = \sum_j\beta_j(t)\psi_j(s)+\varepsilon(s,t)
\end{equation}
for some basis functions $\psi_j$. For the first three models proposed, $\varepsilon(s,t)$ is a spatiotemporal white noise, however the FSBGL models this using a compactly supported covariance kernel in space. The resulting spatial covariance matrices of the fields can be decomposed as
\begin{equation}
    \label{eq:frk_eqn}
    \mathbf{\Sigma} = \mathbf{\Psi}\mathbf{Q}^{-1}\mathbf{\Psi}^T+\mathbf{D}
\end{equation}
with $\mathbf{Q}$ the precision matrix of the coefficients $\beta_j$.
In the event that $\m{\Psi}$ has a unique QR decomposition, estimates of the coefficient covariance matrix $\hat{\m{Q}}^{-1}$ and nugget variance $\hat{\tau}^2$ can be found in closed form by minimizing the Frobenius norm of the differece between $\mathbf{\Sigma}$ and the sample covariance matrix \citep{frk}. However, since the solid needlets form a frame of a polynomial space rather than a basis, they are collinear and the matrix $\m{\Psi}$ does not have full rank. Thus, the QR decomposition of $\m{\Psi}$ is not unique, making the closed form solution provided in \cite{frk} unusable. To avoid this, for the FRK model we replace $\m{\Psi}$ with the matrix $\m{U}$ whose columns are the left singular vectors of $\m{\Psi}$ with nonzero singular values. Due to the dense structure of the matrices involved, FRK is computationally intractable on large datasets. For LK, the basis functions are Wendland kernels \citep{wendlandrbf} and the precision matrix is parameterized via a spatial autoregression with the parameters learned from the data \citep{nychka}. This results in an efficient and flexible model, however it currently generates stationary covariance kernels and does not allow for cross-correlation between levels of resolution. This limitation motivates the BGL and FSBGL, which both estimate $\m{Q}$ by means of a lasso penalty \citep{bgl,fsbgl}, with the models differing in the form of $\m{D}$. The BGL assumes $\m{D}=\tau^2\m{I}$, while the FSBGL assumes it has a compactly supported covariance. Both of these models allow more expressive structure in the low-rank portion of the covariance while retaining computational tractability by heavily leveraging sparsity.

Applying the FSBGL comes with a choice about which small-scale covariance model to use. We will consider three different covariance models for $\varepsilon(s,t)$ based on a Wendland kernel which vary slightly in the assumed geometry. Due to the advection in the radial direction by winds, we expect to see that the correlation length in the tangential direction is different from the radial direction. One option would be to use a polar Gaussian process, wherein separate covariance kernels are used in the radial and tangential direction (\cite{padonou2016polar}), however this results in a stationary kernel that preliminary analysis suggests is not appropriate for the small-scale model. We choose a slightly different approach via a conformal mapping from the unit ball to the upper half of the unit sphere given by
\begin{equation}
    (r,\theta) \mapsto (2\arctan(r),\theta)
\end{equation}
we can then embed the unit ball in $\mathbb{R}^3$ as
\begin{equation}
\vec{z}(r,\theta)
=
\left(
\sin(\phi)\cos(\theta),
\sin(\phi)\sin(\theta),
\cos(\phi)
\right)^\top,
\qquad \phi=2\arctan(r),
\end{equation}
and define an anisotropic distance metric by
\begin{equation}
d_{\boldsymbol{\ell}}(\vec{z}_i,\vec{z}_j)
=
\left[
(\vec{z}_i-\vec{z}_j)^\top
\boldsymbol{\Lambda}^{-1}
(\vec{z}_i-\vec{z}j)
\right]^{1/2},
\qquad
\boldsymbol{\Lambda}
=
\operatorname{diag}(\ell_1^2,\ell_2^2,\ell_3^2).
\end{equation}
This guarantees a positive definite covariance kernel and naturally allows the correlation lengths to vary in the $r$ and $\theta$ directions. We consider both a two-parameter version, in which $\ell_1=\ell_2$, and a three-parameter version in which all three correlation scales are estimated separately. The former preserves rotational symmetry about the storm center, while the latter allows for additional storm-relative directional dependence. These models are, respectively, called FSBGL-EW2 and FSBGL-EW3. We compare them to a simpler Wendland model without projection onto the unit sphere where the distance metric is given by a Euclidean distance in $\mathbb{R}^2$, with correlation lengths varying in the $x$ and $y$ directions. This model will be referred to as FSBGL-W. After simulation, the portions over land are smoothed with a Gaussian filter to better represent the behavior over land.

\section{Results}\label{sec:verify}

This section describes the results and model validation procedures. First we will examine some draws from the model as compared with the WRF data and Holland model in Section \ref{sec:sim_fields}, then examine the bulk flow and small-scale models in Sections \ref{sec:validate_bulk} and \ref{sec:validate_resids}. We will specifically examine how well the model captures variability in the kinetic energy along the coastline in Section \ref{Sec:coastline_ke} and model calibration in Section \ref{sec:model_cal}.

\subsection{Simulated Fields}\label{sec:sim_fields}

\begin{figure}
    \centering
    \includegraphics[width=\linewidth]{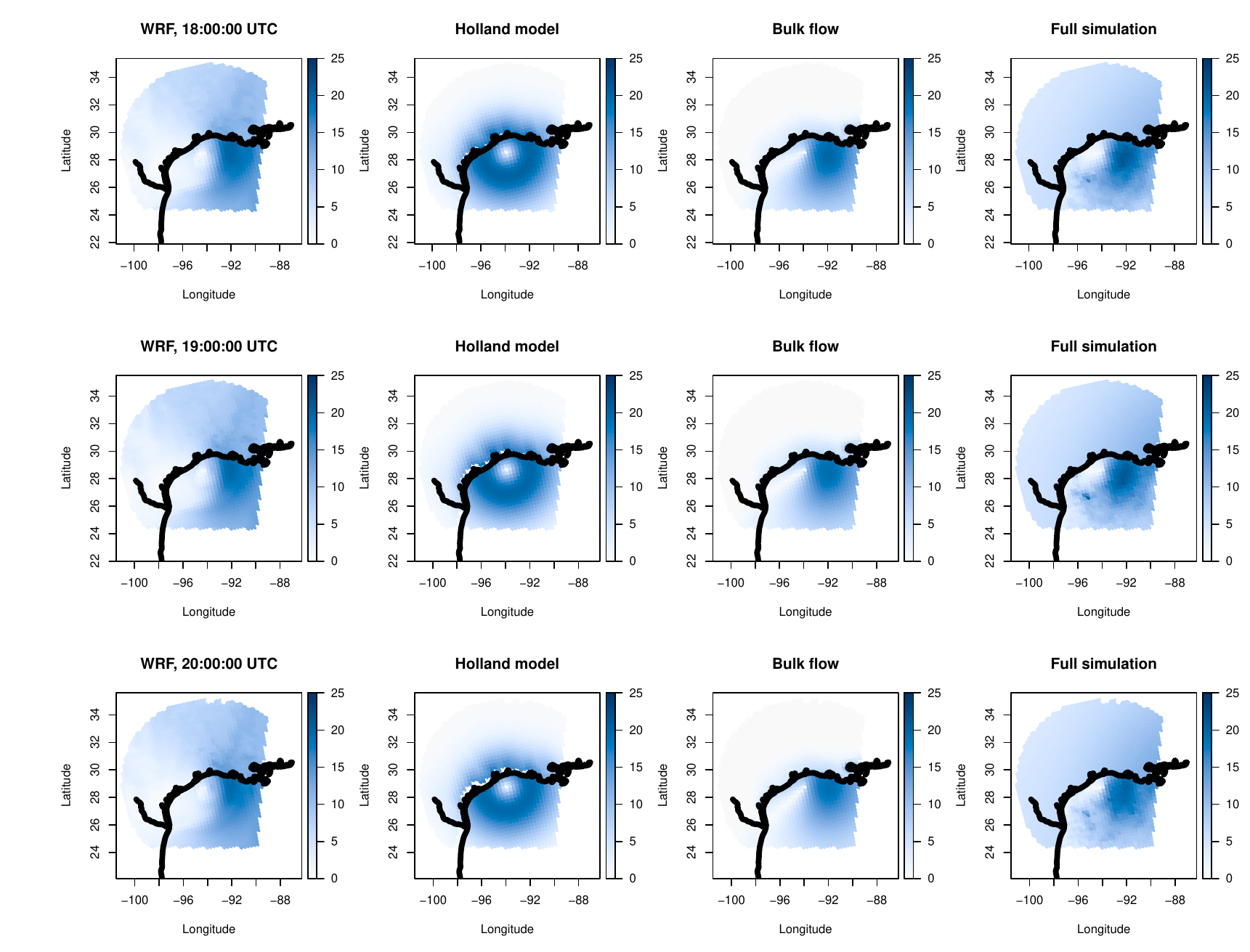}
    \caption{WRF data (left), the Holland model (left-center), a draw from the bulk flow model (right-center) and a draw from the full stochastic wind field model (right) for three consecutive hours corresponding to Tropical Storm Claudette (1979). Wind speeds are in m/s.}
    \label{fig:model_components}
\end{figure}

Before examining the model performance, we first want to show how they combine to create realizations of the wind field. Figure \ref{fig:model_components} compares WRF wind fields with the corresponding Holland radial profile, the fitted bulk flow model, and realizations from the full stochastic model at three representative time steps. The Holland model is able to match roughly the radial shape along the direction of maximal winds, however it is a symmetric vortex which does not represent the shape of real TCs. The addition of the asymmetry in the bulk flow model allows the simulated storms to take a more realistic shape and to evolve stochastically, however the localized behavior of the wind field is not captured, and the model's radial and tangential dynamics are completely separate. The small-scale model corrects both by generating stochastic small-scale structure via a nonstationary covariance kernel, allowing the structure of the perturbations to vary over the storm in a coherent but non-separable manner. This allows the model to be more expressive while still generating realistic wind fields. Lastly, along with the small-scale model we apply a Gaussian smoother to the data over land to generate slightly more realistic wind fields. 

\subsection{Validation of the Bulk Flow Model}\label{sec:validate_bulk}

In this section we address the performance of the bulk-flow model described in Section \ref{sec:mean_model_tcs}. We begin by describing the process for selecting the penalty, then examine the components of the model and the dynamics of the corresponding wind field. 

\subsubsection{Penalty Selection}

Estimating the impulse response function is done via the augmented least-squares problem in Eq. \eqref{eq:irf_least_squares}. The impulse response estimation is regularized using a quadratic penalty of the form
\begin{equation}
    \label{eq:regularizer_irf}
    \mathbf{L}^T\mathbf{L}=\lambda_r\mathbf{I}+\lambda_s\mathbf{P}_s
\end{equation}
where $\mathbf{I}$ is the identity matrix and $\mathbf{P}_s$ is the penalty associated with the different penalties, namely penalties on the first and second time derivative and the first spectral derivative.The form of the penalty and the values of $(\lambda_r,\lambda_s)$ were selected to minimize the AIC given by
\begin{equation}
    AIC = -2\ell(\mathbf{G}) +2df_E
\end{equation}
where $df_E=\tr{\mathbf{H}(\mathbf{H}^T\mathbf{H}+\mathbf{L}^T\mathbf{L})^{-1}\mathbf{H}}$
is the effective degrees of freedom of the estimator and $\mathbf{H}$ is the matrix of observed innovations used in Eq. \eqref{eq:irf_least_squares}. The final penalty selected was the penalty on the first spectral derivative, although the model selected a relatively modest weight of $\lambda_s=10^{-7}$ and a larger weight of $\lambda_r=10^{-4}$ for the ridge penalty. These are the parameters used for the model validation. 

\subsubsection{Random Effects}

The first component of the model for the Fourier coefficients describing the angular variation is a stormwise random intercept term $c_{i,j}$. This component of the model, which makes up a significant portion of the variability, is meant to capture variability from storm to storm not captured by the predictors, for example varying sea surface temperature, wind shear, or solar heating. Further, draws from the random effects distribution are used in simulation and validation of this model. For this reason, we want to check whether the fitted random effects are physically meaningful and describe a coherent angular structure in the vector field. To do this, we performed principal component analysis of the fitted random effect vectors for each storm and calculated the angular wind profiles in the cross-track and along-track direction induced by the first four principal components. 
This number was chosen because they explain 81\% of the variance, and because given our sample size of 19 storms restricting the model to four components satisfies traditional sampling adequacy constraints (e.g. \cite{cattelfactoranalysis}) while ensuring stable component recovery due to the high variance explained \citep{maccallum1999sample,mundfromminimalpca}. However, due to the relatively low sample support we damp the fitted random effect variances by a factor of 0.8.
\begin{figure}[h!]
    \centering
    \includegraphics[width=0.95\linewidth]{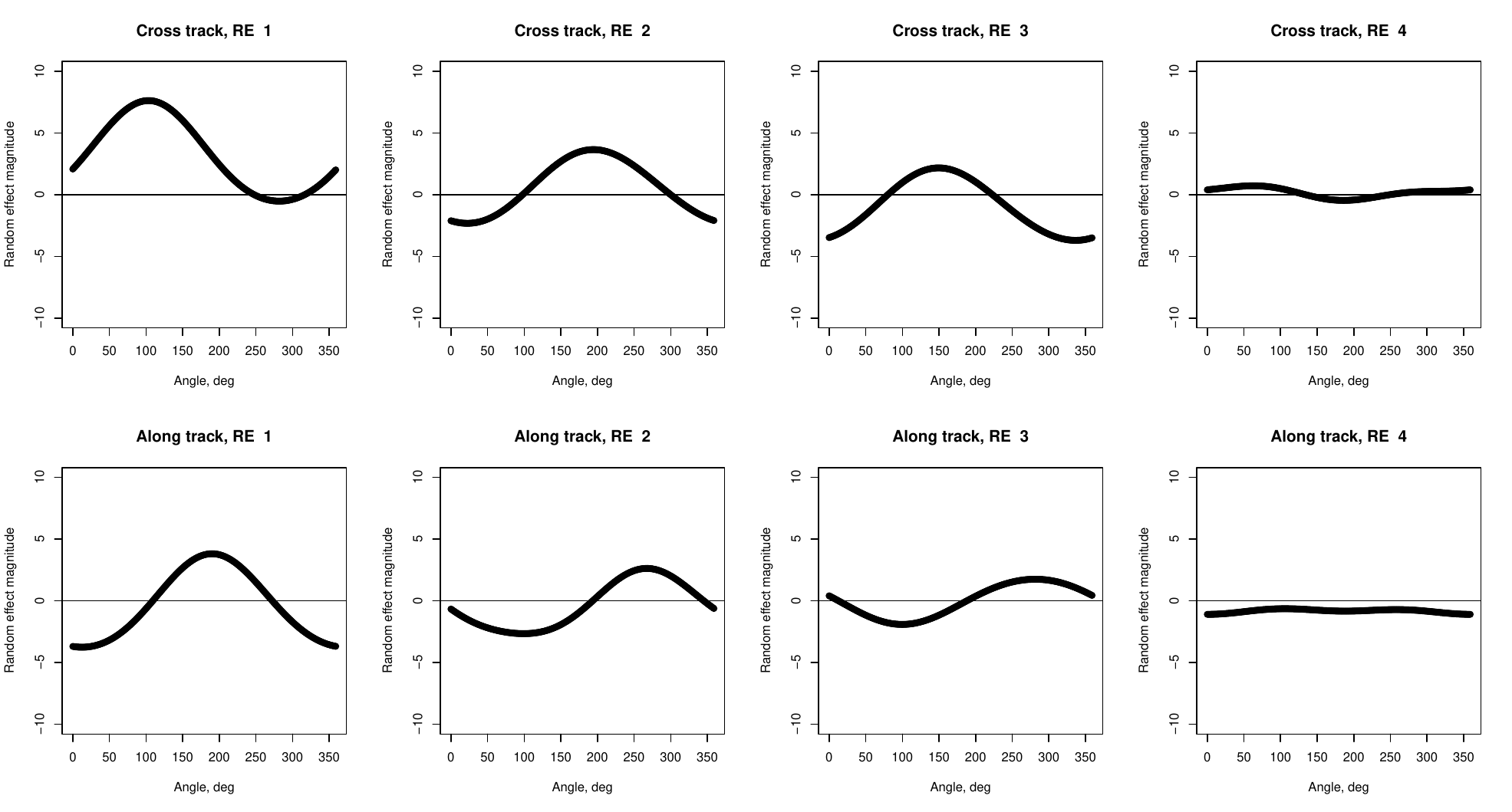}
    \caption{Angular wind profiles corresponding to the first four eigenvectors of the random effects covariance scaled by the principal component standard deviation.}
    \label{fig:random_effects}
\end{figure}

The cross track and along track directions of the first 4 principal components scaled by their standard deviations are plotted in Figure \ref{fig:random_effects}. The random effects show physical behavior, mostly describing slight perturbations to wavenumber-1 structure or serving to shift the angle of maximum winds. Due to the aforementioned effects of variables such as wind shear on the storm shape, these appear physically reasonable as modes of variation.  All validation in the following sections will be done with simulated random effects. 

\subsubsection{Stochastic Evolution of the Fourier Coefficients}\label{sec:stochastic_evolution_irf}

The model for the stochastic residuals on the Fourier coefficients is complex, and it is natural to wonder whether that additional complexity is necessary. 
To investigate this, we fit a VAR(1) process to the data and investigated the behavior of the fitted model relative to the data. 
\begin{figure}
    \centering
    \includegraphics[width=0.85\linewidth]{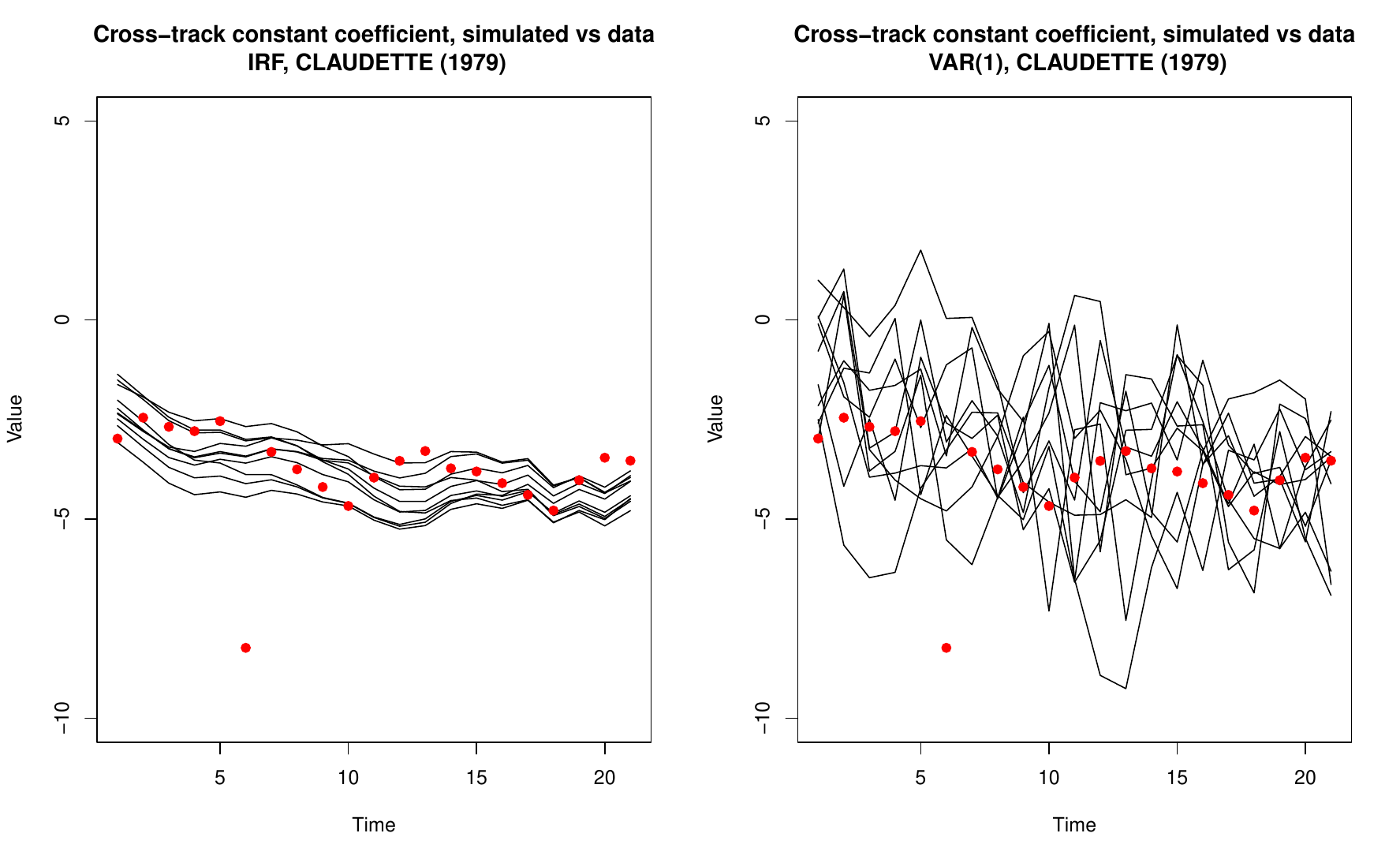}
    \caption{Simulations using the IRF model (left) and a VAR(1) fit to the residuals (right). The VAR(1) is far too rough to accurately fit the dynamics observed in the data.}
    \label{fig:cc0_irf_v_var}
\end{figure}

For the model to appear realistic, simulated coefficient trajectories containing both the mean and the residuals need to be able to match the coefficient trajectories from the data. The coefficient trajectories in the data are quite a bit smoother than implied by the observed residuals, and so we need to ensure that the simulated trajectories are also sufficiently smooth. Unfortunately, the innovations of a VAR(1) process are too rough for this, as seen in Figure \ref{fig:cc0_irf_v_var}. Both simulations were done using the fitted random effects for Claudette to emphasize the difference in the behavior of the sample paths. This motivates the more complex IRF formulation, which is able to generate smoother, less volatile sample paths that more closely resemble the behavior of the data.

\begin{table}[h!]
    \centering
    \begin{tabular}{|c|c|c|c|}
    \hline
      Parameter & $\sigma^2$ & $\kappa^2$ & $\nu$  \\
        \hline
        Value & $4.7\times 10^{-3}$ &  0.062 & 2.85\\
        \hline
    \end{tabular}
    \caption{Fitted Mat\'ern parameters for the spectrum of the common innovation dynamics.}\label{tab:coeff_matern_params}
\end{table}

\begin{figure}[h!]
    \centering
    \includegraphics[width=0.8\linewidth]{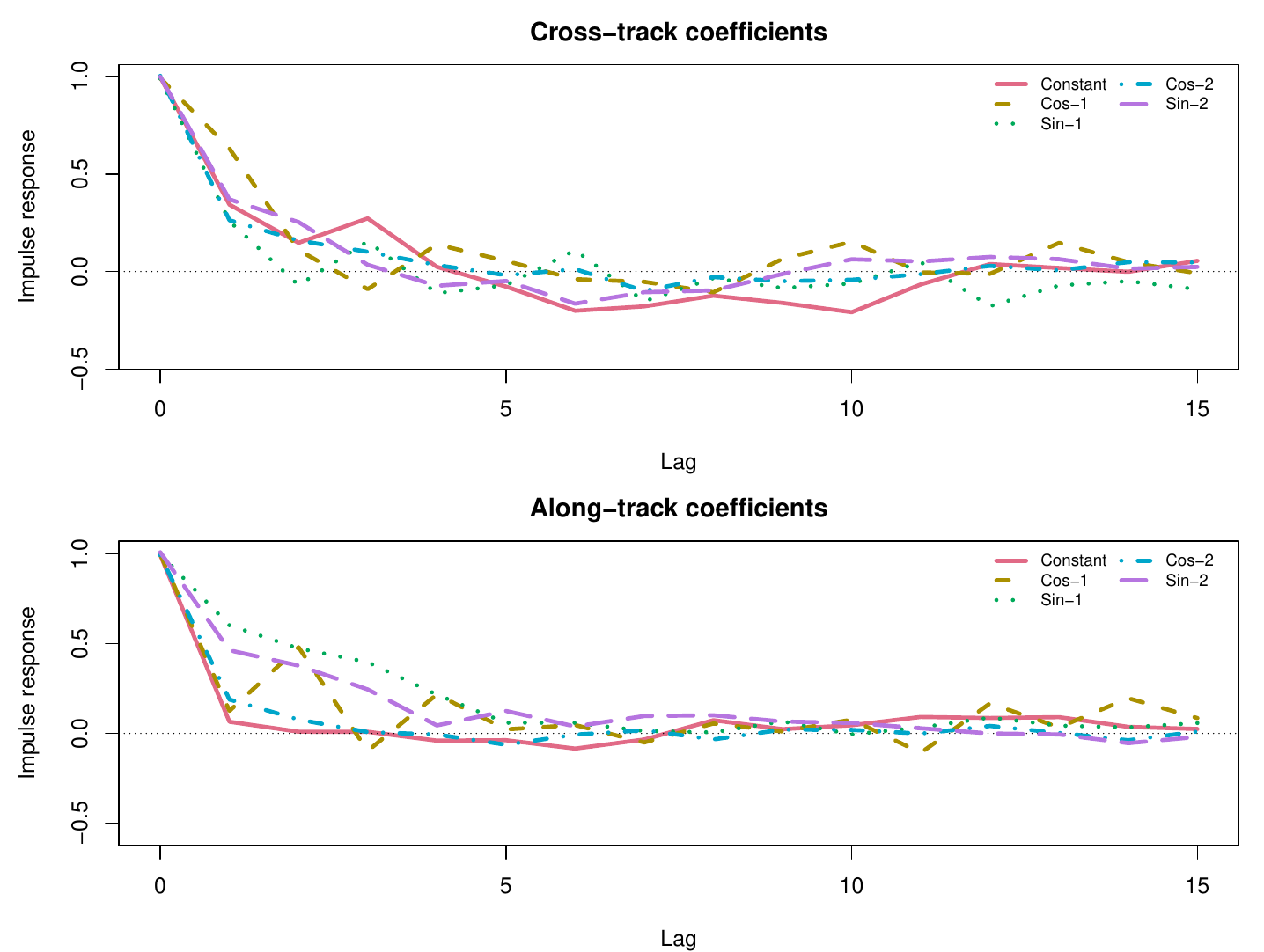}
    \caption{Fitted impulse response functions for defining lag-dependent dynamics that vary between coefficients. Most coefficients are similar dynamically with the exception that the along track sine coefficients show much longer decay times than the rest. }
    \label{fig:irf_plot}
\end{figure}

In our model, the dynamics are induced jointly by the spectrum, which describes common dynamics, and the impulse response which defines lag-dependent corrections. To avoid overfitting on a relatively limited dataset, we fit a Mat\'ern power spectrum to the observed spectrum. The fitted Mat\'ern parameters are shown in Table \ref{tab:coeff_matern_params}. The fitted smoothness parameter is $2.85$, suggesting differentiable sample paths for the coefficients. The diagonals of the impulse response function for lags 0-15 are plotted in Figure \ref{fig:irf_plot}. The cross-track coefficients generally exhibit oscillatory decay, with several responses becoming negative at intermediate and longer lags before returning toward zero. The along-track sine terms demonstrate the most persistent dynamics of any modes, while the constant and cosine modes decay extremely rapidly. Since the wavenumber-1 sine mode is best aligned with the expected structure of the along-track winds, this demonstrates the relative stability of the behavior in that direction. The relative persistence of the wavenumber-2 mode suggests that the deviation in the along-track direction from wavenumber-1 structure is also fairly persistent. 

\subsubsection{Evolution of the Angle of Maximum Winds}
One of the most important structural components of the wind field is the angle of maximum winds. This angle measures not only the departure of the storm from expected angular structure but also determines the location relative to the storm where winds are greatest and thus which locations are most at risk of severe damage from extreme winds \citep{olfateh2017asymmetry}. In addition, it is one of the physical features of the storm which the model is not trained on due to it not appearing in track databases and is important to model well for simulation realism. For these reasons, it makes a natural test parameter for our model.

\begin{figure}[t!]
    \centering
    \begin{subfigure}[b]{0.5\textwidth}
       \includegraphics[width=0.85\linewidth]{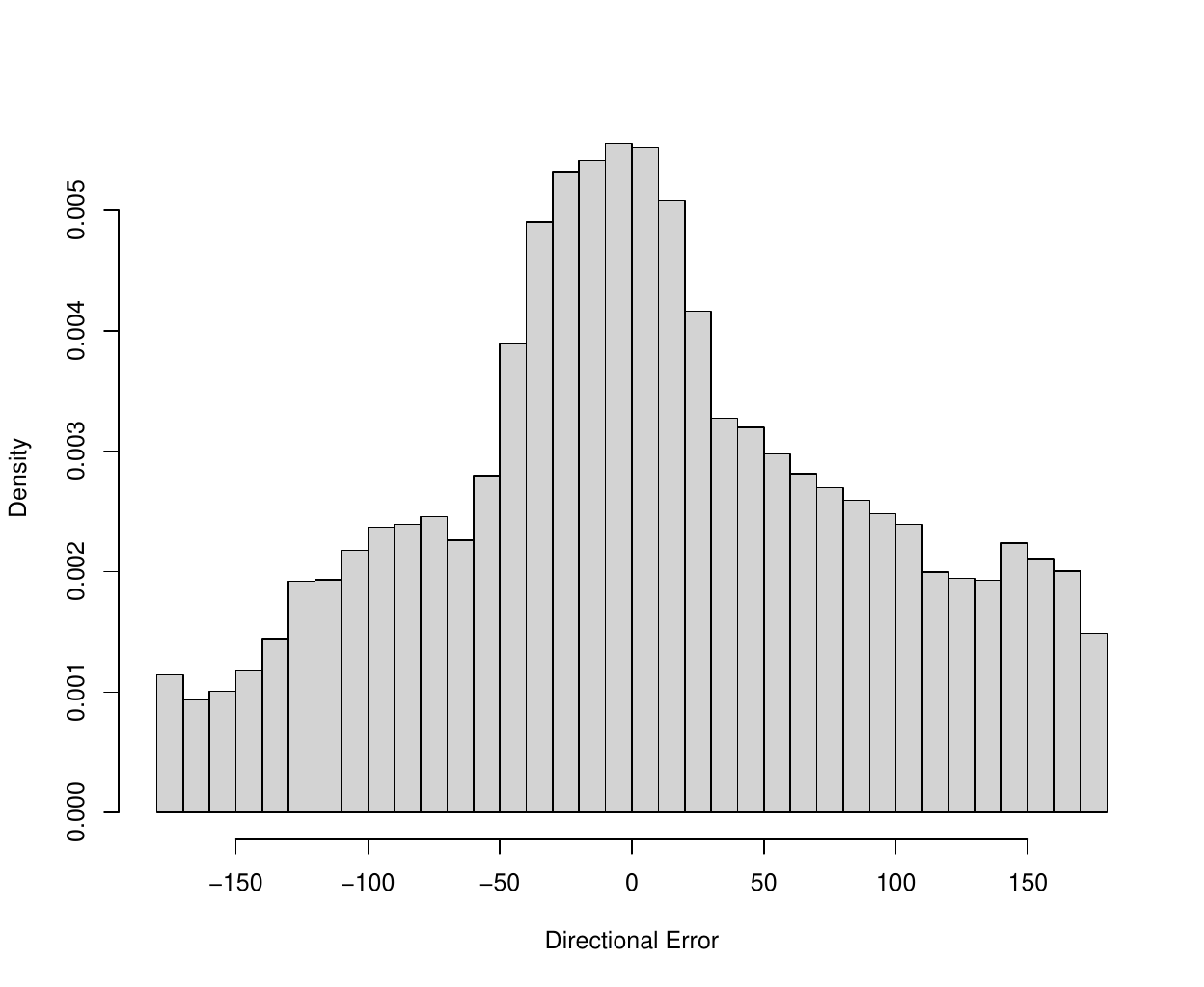}
     \caption{Directional error}\label{fig:de_histogram}
    \end{subfigure}%
    \begin{subfigure}[b]{0.5\textwidth}
           \includegraphics[width=0.85\linewidth]{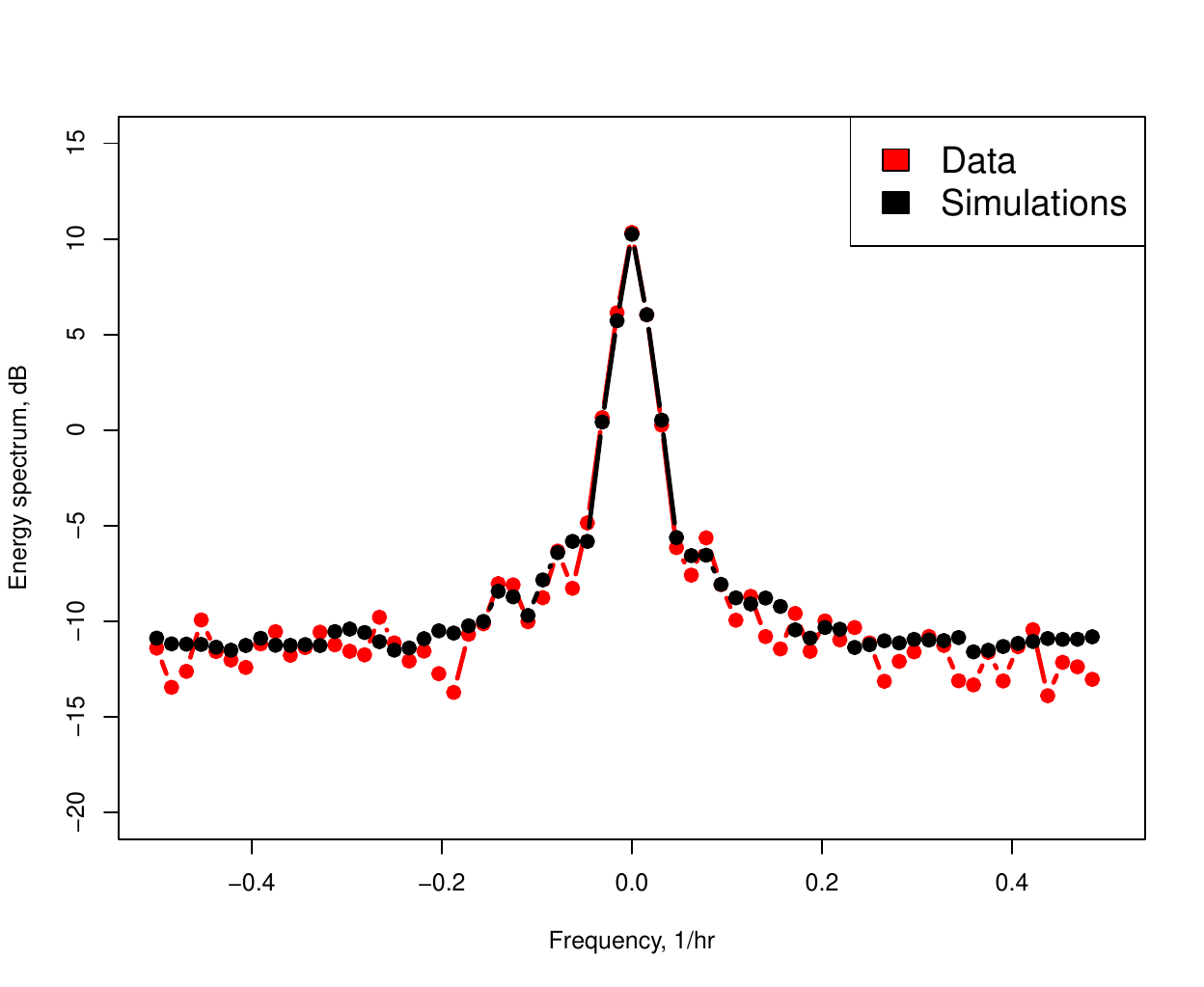}
     \caption{Spectrum of Angular innovations} \label{fig:de_spectrum}
    \end{subfigure}
    \caption{Histogram of directional error in cross-validation (left) and energy spectrum of angular innovations (right). Black denotes the spectrum from simulation and red from the data.}
    \label{fig:de_results}
\end{figure}

In order to investigate the behavior of the angle of maximum winds we first want to examine the angular difference between the angle observed in the data and the simulated angles. To do this, we performed leave-one-out cross-validation on each storm in the dataset and generated 750 realizations of the wind field for each storm. The directional error was calculated as \citep{mardia2009directional}
\begin{equation}
    \text{DE} = \text{atan2}\left(\sin \left(\hat{\phi}-\phi_s \right), \cos \left(\hat{\phi}-\phi_s\right)\right)
\end{equation}
where $\hat{\phi}$ is the angle from the data and $\phi_{s}$ is the simulated angle. A histogram of these errors on the training data and from cross-validation is shown in Figure \ref{fig:de_histogram}. The model is generally able to predict the angle of maximum winds well based on these histograms, although not perfectly. This is not unexpected, as this angle is dependent on many environmental variables like wind shear that do not make it into the model, so it will struggle on unobserved data \citep{ObservedHurricaneWindSpeedAsymmetriesandRelationshipstoMotionandEnvironmentalShear}. Additionally, for weaker storms this angle may be more poorly defined due to the relative symmetry of the storm, and so the model may struggle to match that behavior. 

Another important component of simulation realism is the ability to capture the temporal behavior of different features of the model. For this reason, we want to examine the variability of the innovations $\Delta \phi_n = \phi_n-\phi_{n-1}$
in the phase angle. However, this is an angular quantity and so changes from, for example, $\phi=0\degrees$ to $\phi=357\degrees$, can be misinterpreted as large jumps if care is not taken in their analysis \citep{breckling1989analysis}.
Several options exist for the analysis of autocorrelation of angular data, for example \citep{artes2009autoregressive,fisher_timeseries_circular}. We choose to determine the spectrum of the derived quantity \citep{craig1989time}
\begin{equation}\label{eq:z_innovs_angle}
    z_n = \exp(i\Delta \phi_n)
\end{equation}
which allows us to treat the data as any other time series, albeit a complex-valued one. The energy spectrum of the time series $z_n$ for the data and simulations on cross-validation data are shown in Figure \ref{fig:de_spectrum}. The agreement is excellent, which suggests that the model is able to accurately capture the dynamic evolution of the angle of maximum winds, even for storms that are not in the training data. This combined with the accuracy of the simulated angles of maximum winds suggests that the model is able to generate realistic simulations of storms that were not in the training data. 

\subsubsection{Evolution of Asymmetry}

Much of the recent development of parametric hurricane models, including the model developed here, has been motivated by the desire to properly capture angular asymmetry in the storm structure. One indicator of the wind field asymmetry is given by the asymmetry index \citep{olfateh2017asymmetry}
\begin{equation}
    \label{eq:asym_index}
    \delta = \fraC{V_m}{\tilde{V}}-1 , \quad
    \tilde{V} = \frac{1}{2\pi}\int_0^{2\pi}V(R_m,\theta)\diffd \theta
\end{equation}
which describes the fractional difference between the maximum observed wind speed and the mean wind speed at the radius of maximum winds. 
The asymmetry of the storm has significant effects on the severity of storm surge and other processes like ocean cooling, vertical mixing of the water column, and phytoplankton growth in the wake of the storm \citep{ascmo-10-69-2024,huang2015phytoplanktonTC,xie2011numerical}. For this reason, and for the sake of realism of the simulations, we wish to analyze whether both the values of the asymmetry index and the temporal evolution are realistic as based on the dataset. 

\begin{figure}[h!]
    \centering
    \begin{subfigure}[b]{0.5\textwidth}
       \includegraphics[width=0.85\linewidth]{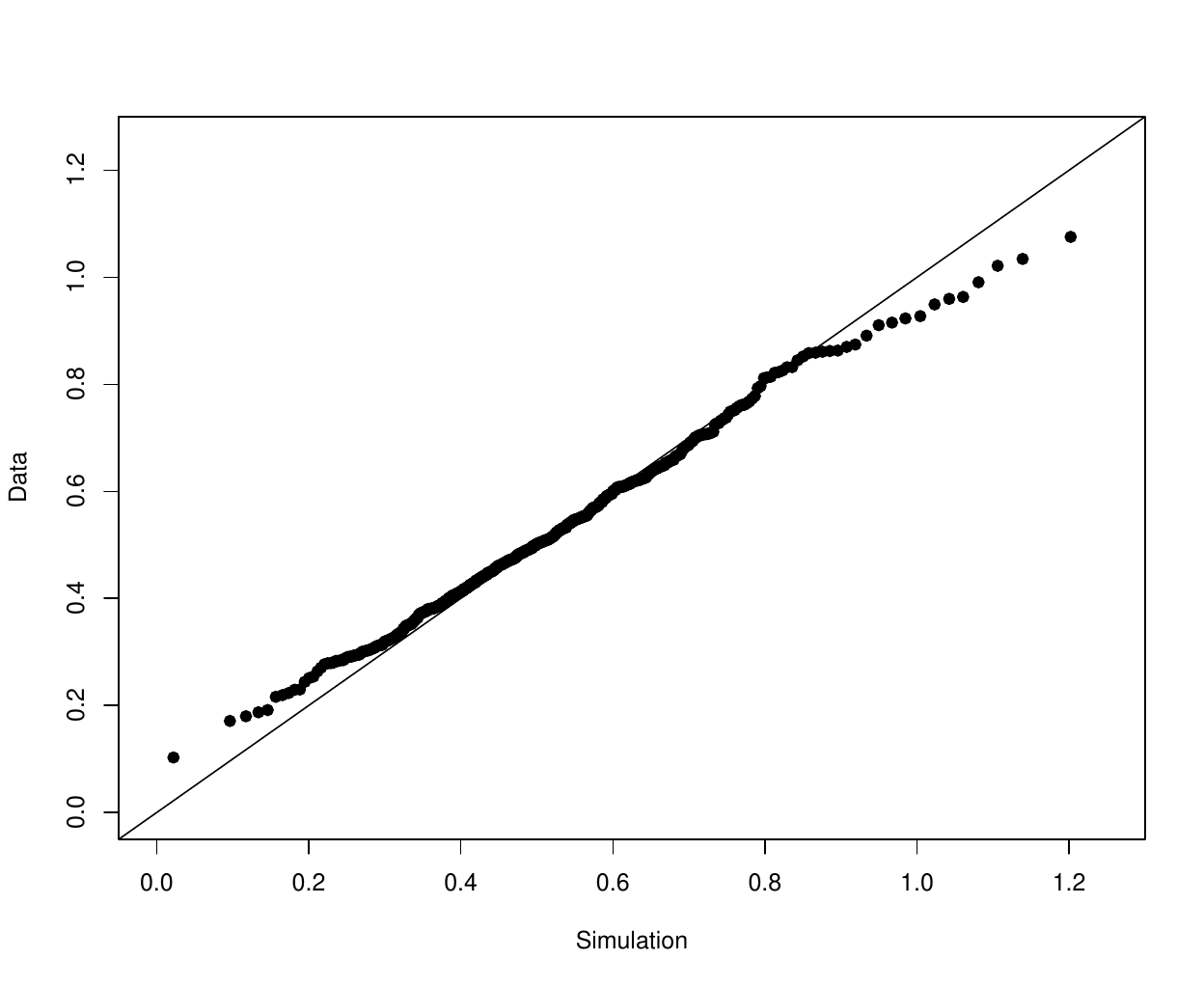}
     \caption{Q-Q plot of asymmetry index}\label{fig:qq_asym}
    \end{subfigure}%
    \begin{subfigure}[b]{0.5\textwidth}
           \includegraphics[width=0.85\linewidth]{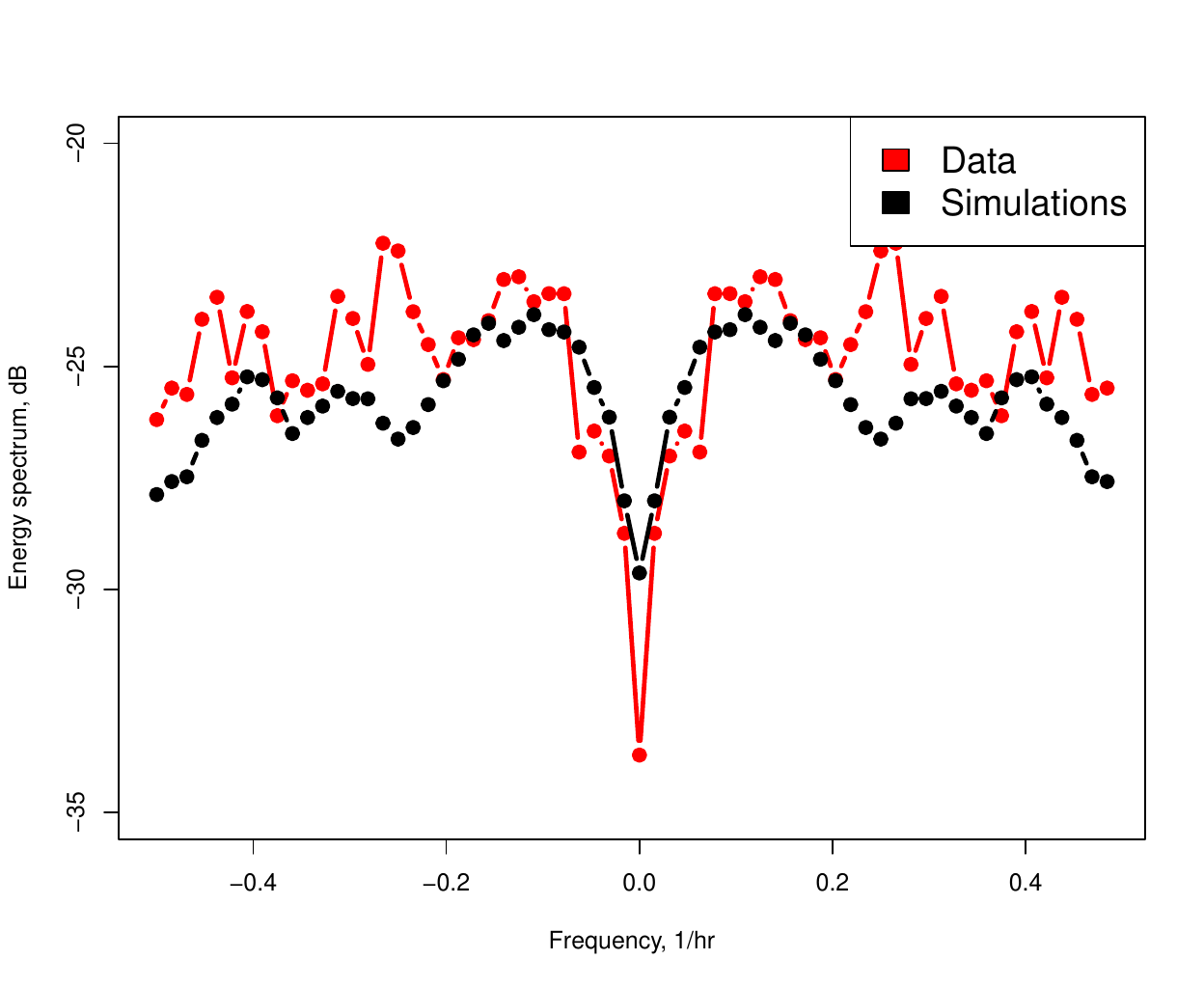}
     \caption{Spectrum of asymmetry index innovations}\label{fig:spectra_asym}
    \end{subfigure}
    \caption{Q-Q plot of asymmetry indices with simulation quantiles along the x axis and data along the y axis (left) and the energy spectrum of asymmetry index innovations as seen in the data (red) and simulations (black).}
\end{figure}

A first investigation requires determining whether the asymmetry indices returned by the model take on realistic values. Due to the complex interaction of the coefficients of the model with the shape of the angular wind profile, it may be possible that a set of physically realistic coefficients yields unrealistic variability in angle.
We investigate this by means of Q-Q plots for the cross-validation dataset used in the previous section, shown in Figure \ref{fig:qq_asym}. The x axis is quantiles of the simulated asymmetry index while the y axis corresponds to quantiles of the asymmetry index derived from WRF. The simulated data captures the bulk of the distribution of the asymmetry index, but the model also generates storms that are both more and less asymmetric than those observed in the data yet remain realistic. This indicates that the stochastic component is capable of producing plausible variability beyond that observed in the finite training sample, rather than simply reproducing the empirical distribution.

Once again, we are also interested in the temporal evolution of the asymmetry index. Figure \ref{fig:spectra_asym} contains plots of the energy spectra of innovations in the asymmetry index from the training set and cross-validation compared to the evolution observed in WRF. We are able to reproduce the shape of the spectra seen in the WRF data, although perhaps with slightly less energy at high frequencies. 


\subsection{Validation of the Small-scale Model}\label{sec:validate_resids}

The small-scale model is fit in three steps. The first step arises when the neural network is fit to learn the land decay. At this point, the model additionally learns the marginal standard deviations at each spatial location as a function of the predictors. 
Once these are learned and the mean model is fit, the log-residuals are standardized by the learned standard deviations and fit using LatticeKrig, FRK, the BGL, and the three FSBGL variants. 
\begin{table}[h!]
    \centering
    \begin{tabular}{|c|c|c|c|}
    \hline
       Model & AIC & EDF & Optimal penalty \\
        \hline
        LatticeKrig &829 &389 &N/A \\
        BGL & -1143 &  105 & 0.00032 \\
        FRK &-1360&47&N/A\\
        FSBGL-W&-1925&32& 0.00178\\
        FSBGL-EW2&-2002&37&0.00178\\
        FSBGL-EW3&-2028&36&0.00178\\
        \hline
    \end{tabular}
    \caption{AIC, effective degrees of freedom, and optimal value of the penalty parameter (if applicable) for each model tested.}
\label{tab:hurricane_residual_model_tests}
\end{table}

\begin{figure}[h!]
    \centering
    \includegraphics[width=0.65\linewidth]{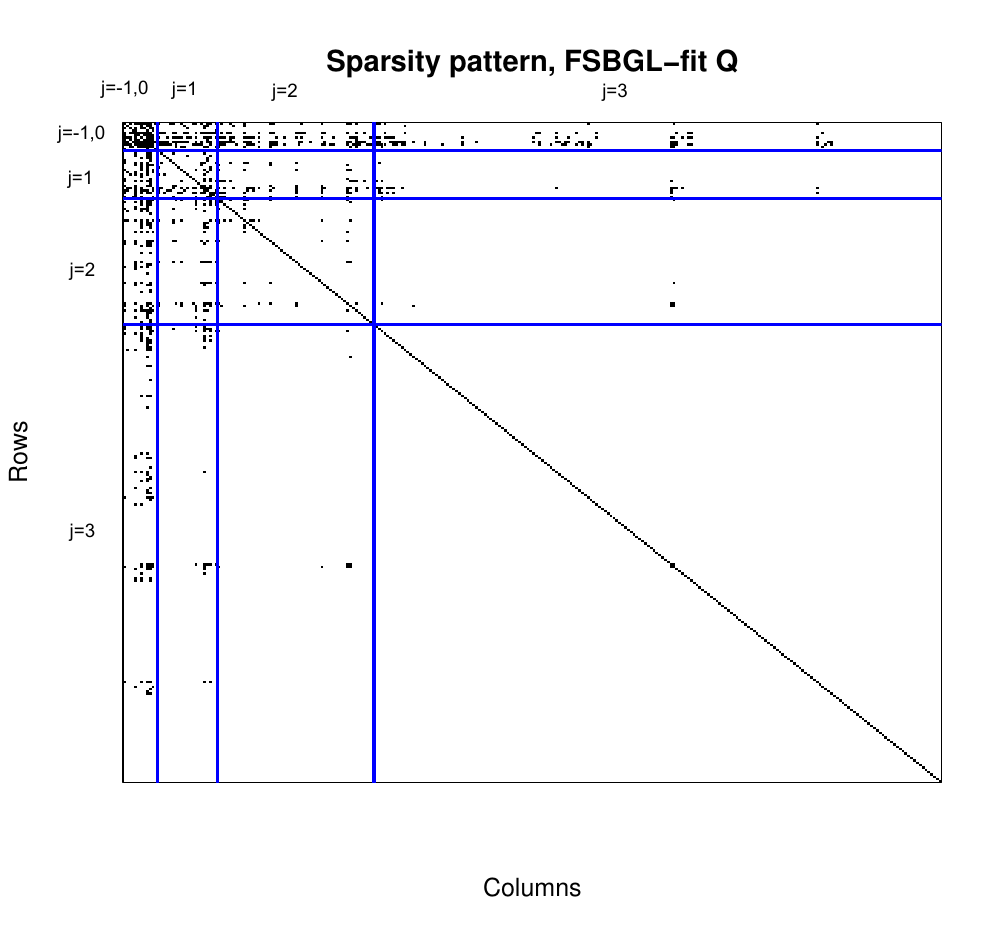}
    \caption{Sparsity pattern in the solid needlet coefficient precision matrix fit by the FSBGL.}
    \label{fig:sneed_Q}
\end{figure}

Model selection, both in terms of the penalty parameter $\lambda$ for the BGL and FSBGL and the overall covariance model, was performed by minimizing the conditional AIC \citep{hatmatrixAIC}. The results are shown in Table \ref{tab:hurricane_residual_model_tests}. The FSBGL outperformed the other models in terms of AIC while also delivering the minimal effective degrees of freedom, even when accounting for the additional parameters needed for the small-scale covariance model. Among the FSBGL results, the two embedded Wendland models saw meaningful improvements in AIC relative to the standard Wendland kernel in spite of a slight increase in the degrees of freedom. We see that the EW3 model is preferred to the EW2 model, suggesting that the additional anisotropy is useful. This may be due to, for example, higher order effects associated with translation such as track curvature or the asymmetric weakening of winds on the left side of the storm. The fitted parameters are shown in Table \ref{tab:hurricane_wendland_params} for the EW3 covariance model, along with a marginal variance $\sigma^2$ and nugget variance $\tau^2$. We see that, while $\ell_1$ and $\ell_2$ are similar, they are both larger than $\ell_3$, as expected. Even though the difference between $\ell_1$ and $\ell_2$ is small, the improvement in AIC of EW3 relative to EW2 suggests that allowing this additional directional flexibility meaningfully improves the fit. Taken together, these results suggest that the dominant anisotropy is between the tangential and radial directions, with a smaller departure from radial symmetry in the tangential direction. 

\begin{table}[h!]
    \centering
    \begin{tabular}{|c|c|c|c|c|c|}
    \hline
      Parameter & $\sigma^2$ & $\tau^2$ & $\ell_1$ &$\ell_2$&$\ell_3$ \\
        \hline
        Value & 0.275 &  0.067 & 0.514& 0.543&0.305\\
        \hline
    \end{tabular}
    \caption{Fitted parameters for the small-scale covariance model.}\label{tab:hurricane_wendland_params}
\end{table}

\begin{figure}[t!]
    \centering
    \includegraphics[width=0.85\linewidth]{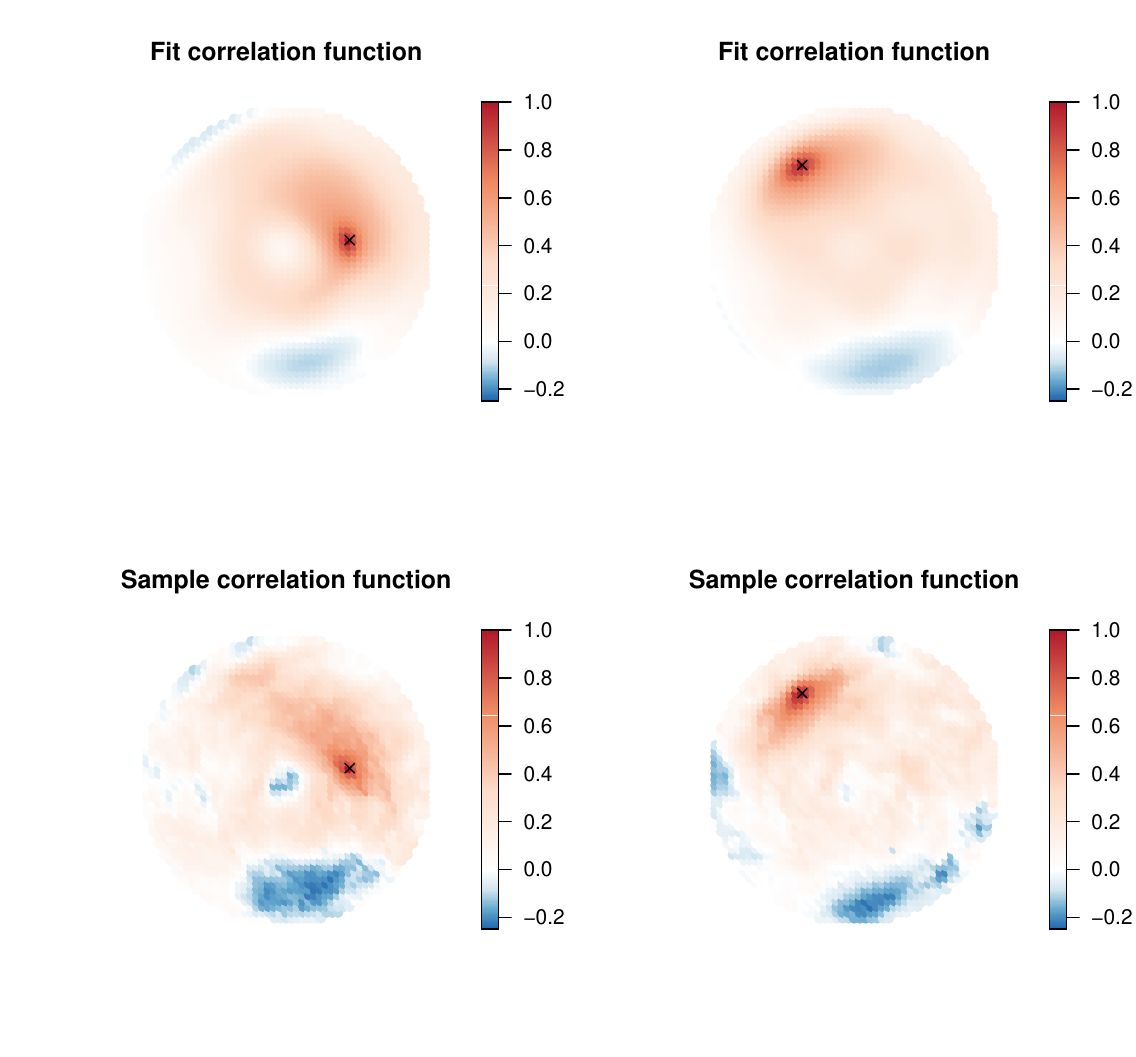}
    \caption{Fit and sample correlation functions of the log-residuals after removing the mean, plotted in storm-relative coordinates. The peak of the correlation function is marked by an "x".
    }
    \label{fig:tc_residual_correlation}
\end{figure}

The fitted sparsity pattern in the precision is shown in Figure \ref{fig:sneed_Q} with the boundaries between resolution levels noted. While the wavelets at the $j=3$ level are mostly conditionally independent, the model has found that they are not conditionally independent of the larger scale structure in the other levels of resolution. In particular, the $j=-1,0$ block appears to play a central role in the cross-resolution conditional dependence structure. This provides a motivation to use methods such as this, as models like LatticeKrig do not allow for this sort of cross-resolution dependence. In Figure \ref{fig:tc_residual_correlation} we can see the sample and fitted correlation functions of the log-residuals centered at two locations. The FSBGL is able to match the large-scale structure well. Notably, the sample correlation functions show weak correlation at the eye of the storm and stronger correlation around the eyewall, which is a feature that the FSBGL is able to reproduce. The agreement illustrates the ability of the joint multiscale covariance model to capture spatial dependence extending well beyond the local correlation structure represented by the compactly supported component or the large-scale structure captured by the low-rank component alone.

\subsection{Variability Along the Coast}\label{Sec:coastline_ke}

\begin{figure}[h!]
    \centering
    \includegraphics[width=\linewidth]{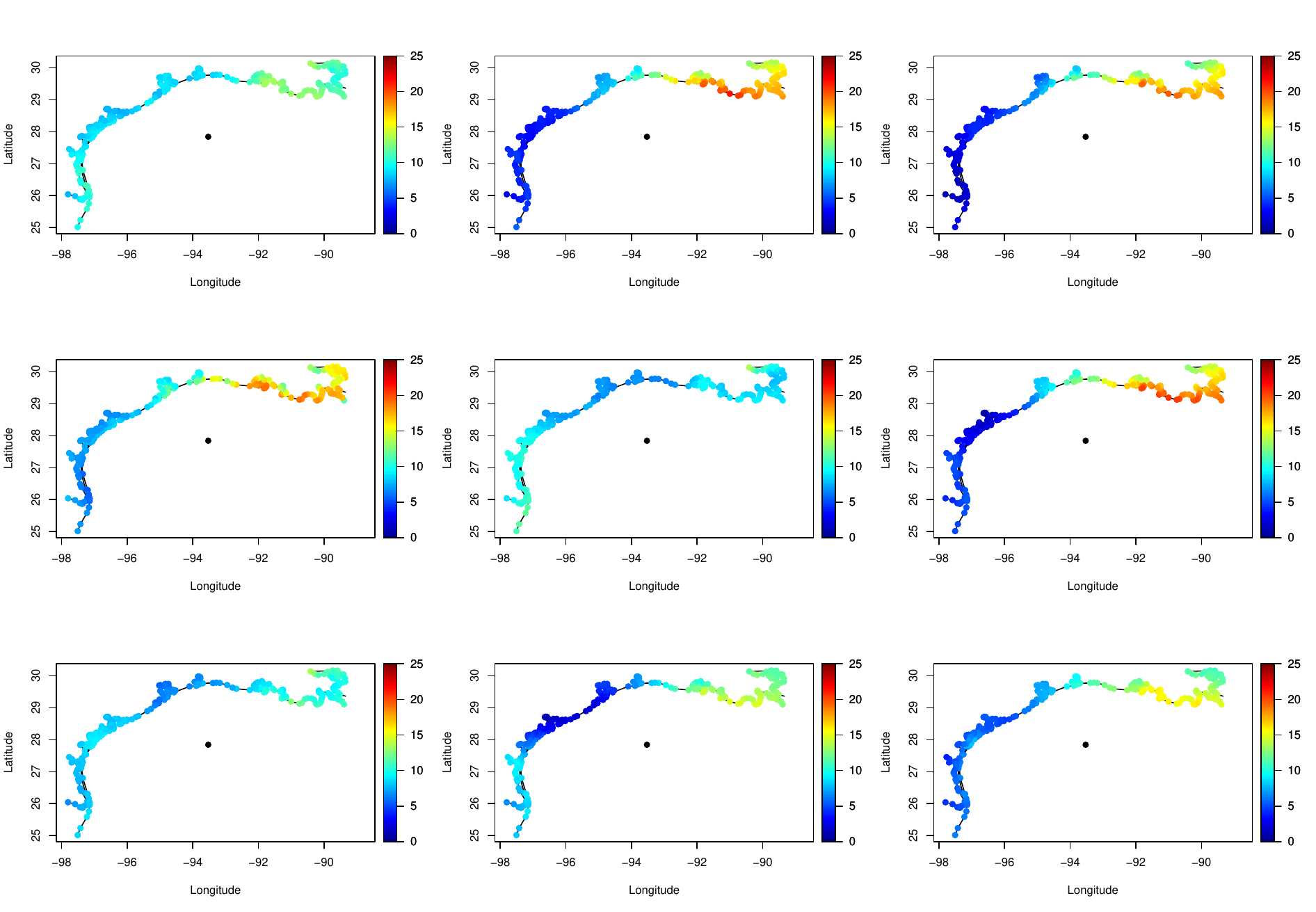}
    \caption{8 simulations of the wind speed in m/s along the coast during Hurricane Claudette and 1 sample from the dataset. The dot denotes the center of the storm, approximately 27.8 deg N by 93.5 deg W. The sample from the training data is on the middle left.}
    \label{fig:sim_v_real_claudette}
\end{figure}

The previous discussion establishes that the model as formulated is capable of reproducing the dynamics of the wind field. However, behavior over specific regions is also relevant to model evaluation and risk assessment. 
Previous work has proposed studying the kinetic energy of TC wind fields as a measure for the potential impact due to winds \citep{TropicalCycloneDestructivePotentialbyIntegratedKineticEnergy}. Motivated by this approach, we examine the kinetic energy of simulated winds along the coastline to assess the potential impacts of landfalling storms. We consider the instantaneous coastline-integrated kinetic energy density (KED) given by 
\begin{equation}
    \label{ke_coastline}
    KED(t) = \frac{\rho}{2}\int_{C}\|\vec{w}(\vec{s},t)\|^2\diffd s\quad  \text{J/m}^2
\end{equation}
where $C$ is the coastline and $\rho\approx 1.225 kg/m^3$ is the density of air. 

Part of modeling this quantity well is accurately reproducing the wind field along the coastline. As a demonstration of our ability to do this, Figure \ref{fig:sim_v_real_claudette}contains 8 randomly selected simulated wind fields and one wind field from the training data, all corresponding to Tropical Storm Claudette (1979). The simulations reproduce the broad spatial structure and range of wind speeds present in the observed field, while exhibiting meaningful and realistic variability between realizations.

\begin{figure}[h!]
    \centering
    \includegraphics[width=0.85\linewidth]{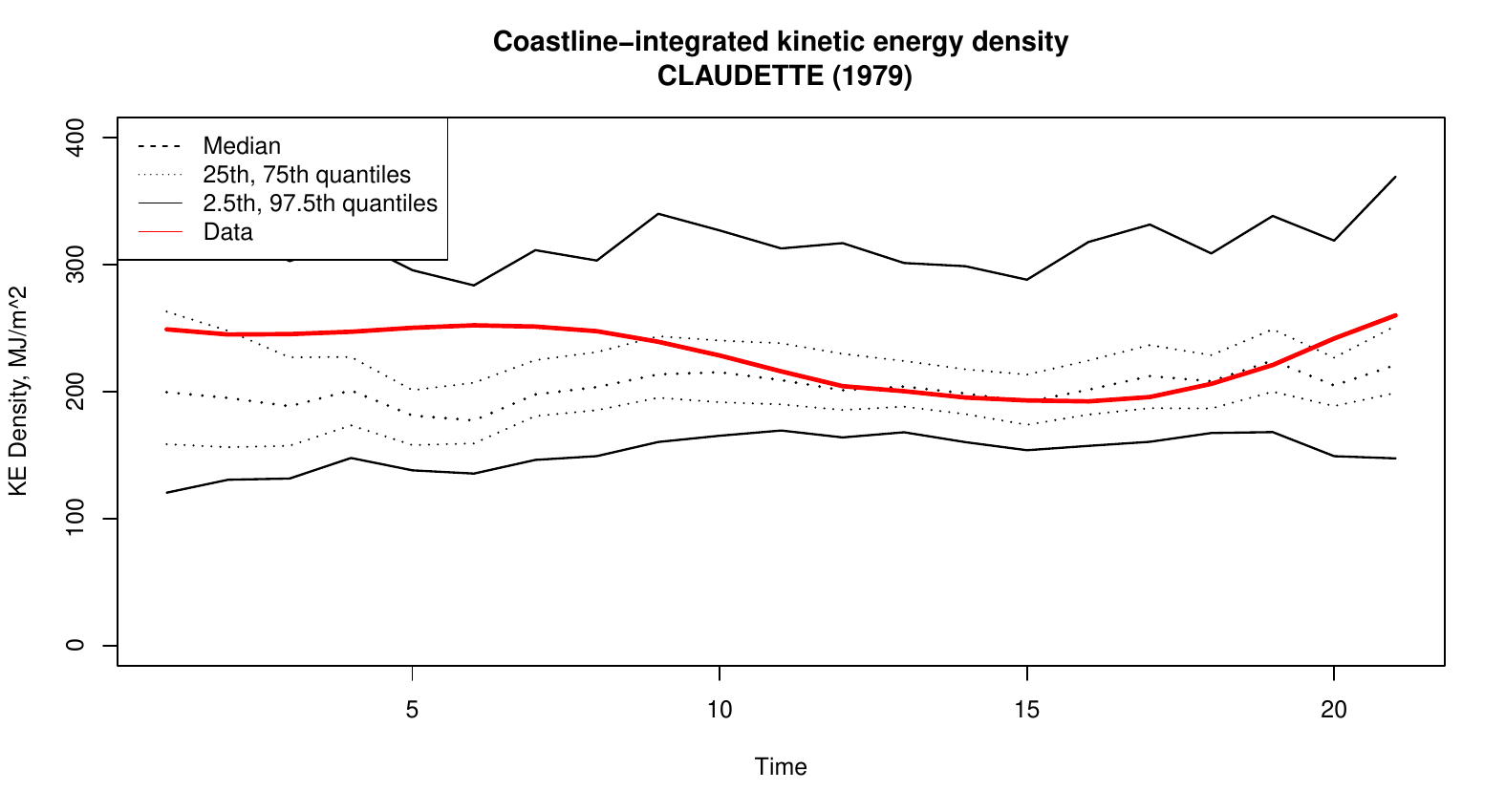}
    \caption{Coastline-integrated kinetic energy density for Tropical Storm Claudette (1979) as a function of time. Results from the WRF run are in red, and the remaining curves represent the mean (dashed) and 2.5th, 25th, 75th, and 97.5th quantiles of the results from the simulation as a function of time. }
    \label{fig:ke_along_coast}
\end{figure}

In order to assess the realism of the model, we calculate the KED for both the WRF winds and 100 simulations of the wind field for Tropical Storm Claudette (1979). The results are shown in Figure \ref{fig:ke_along_coast}. The observed KED generally falls within the central 50\% of the simulated distribution and both the data and simulations exhibit similar temporal evolution. The simulation also demonstrates meaningful skew in the distribution. This suggests that, while the model may slightly underestimate the KED in this instance, it provides reasonable estimates of aggregate wind exposure along the coastline and therefore may be useful in assessing risk derived from this. 

\subsection{Model Calibration}\label{sec:model_cal}

\begin{figure}[h!]
    \centering
    \includegraphics[width=0.85\linewidth]{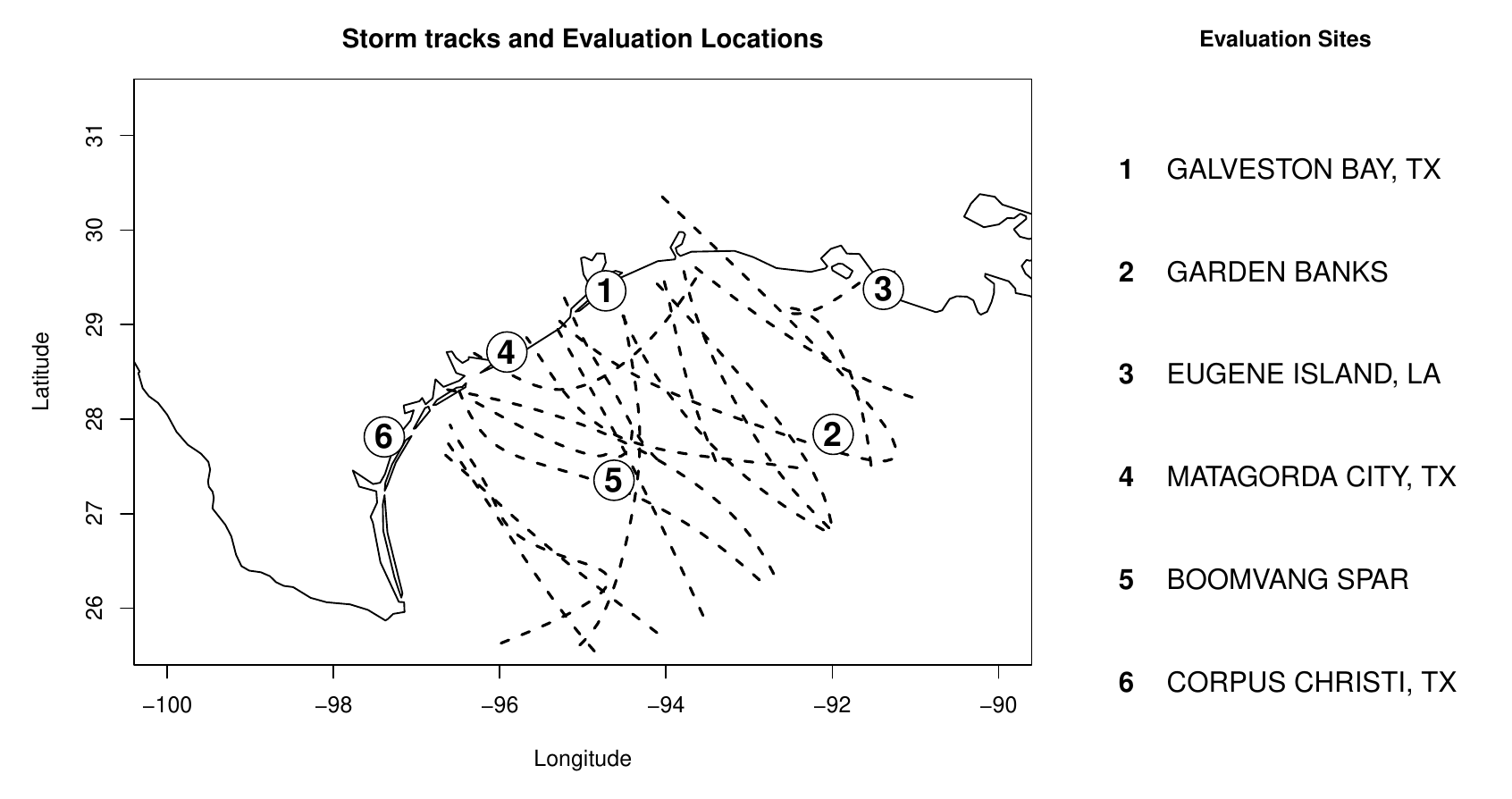}
    \caption{The northwestern coast of the Gulf of Mexico with storm tracks as dashed lines and labeled evaluation locations. }
    \label{fig:storm_tracks_and_eval_locs}
\end{figure}

In addition to the aggregate behavior of the model along the coastline, we are interested in the performance of the model when predicting the wind speed at specific locations, both fixed on Earth and relative to the eye of the storm. In particular, the model must be able to accurately predict winds at cities along the coastline or oil fields in the Gulf of Mexico for it to be useful as a risk assessment tool. These areas are particularly important due to the potential for significant loss of life and economic or environmental damage posed by extreme weather. 

\begin{figure}[t!]
    \centering
    \includegraphics[width=0.85\linewidth]{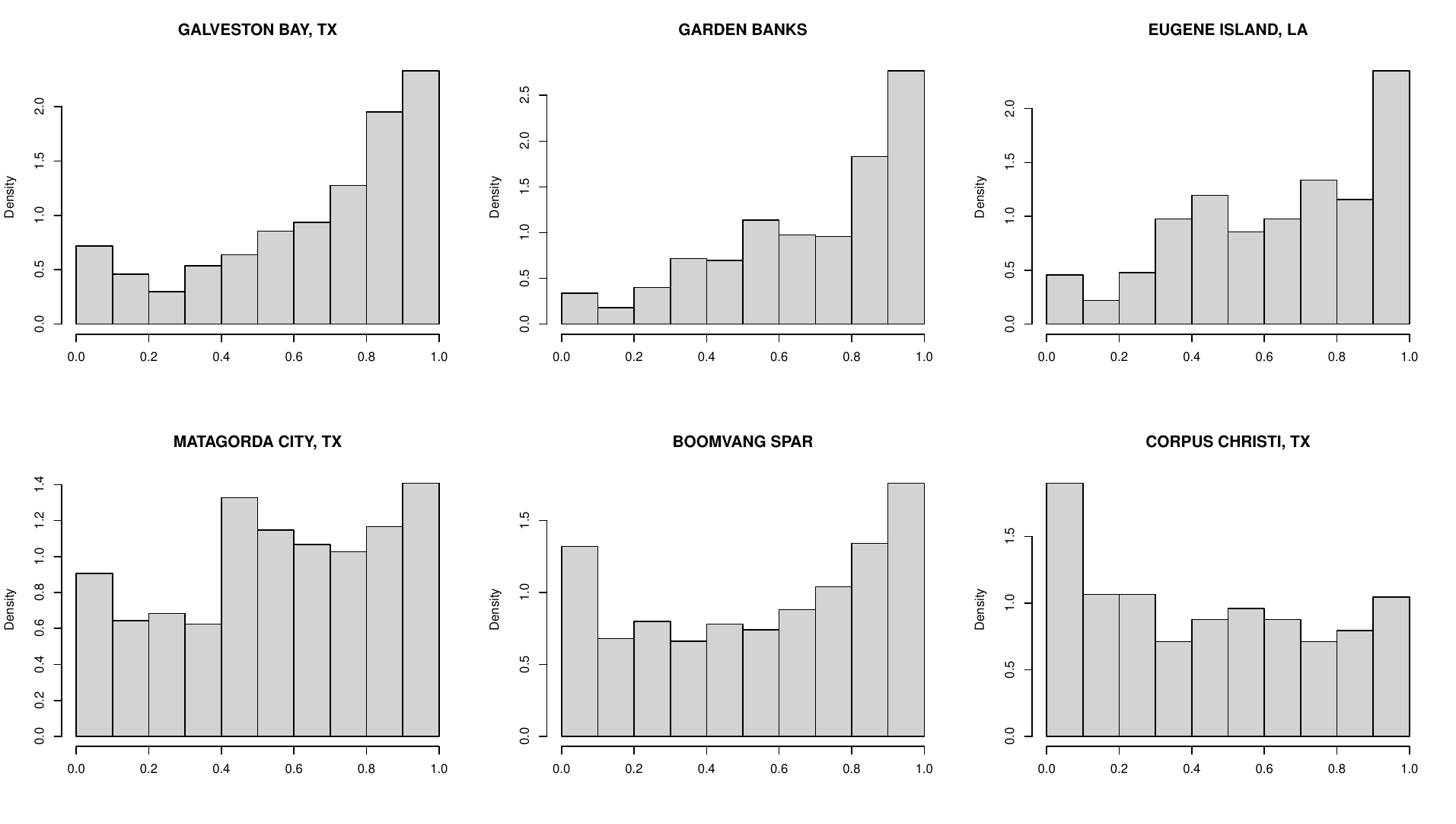}
    \caption{PIT histograms at the 6 geographic locations labeled in Figure \ref{fig:storm_tracks_and_eval_locs}. Calibration results are mixed, likely in part due to missing wind shear effects. }
    \label{fig:space_calibration}
\end{figure}

\begin{figure}[t!]
    \centering
    \includegraphics[width=0.85\linewidth]{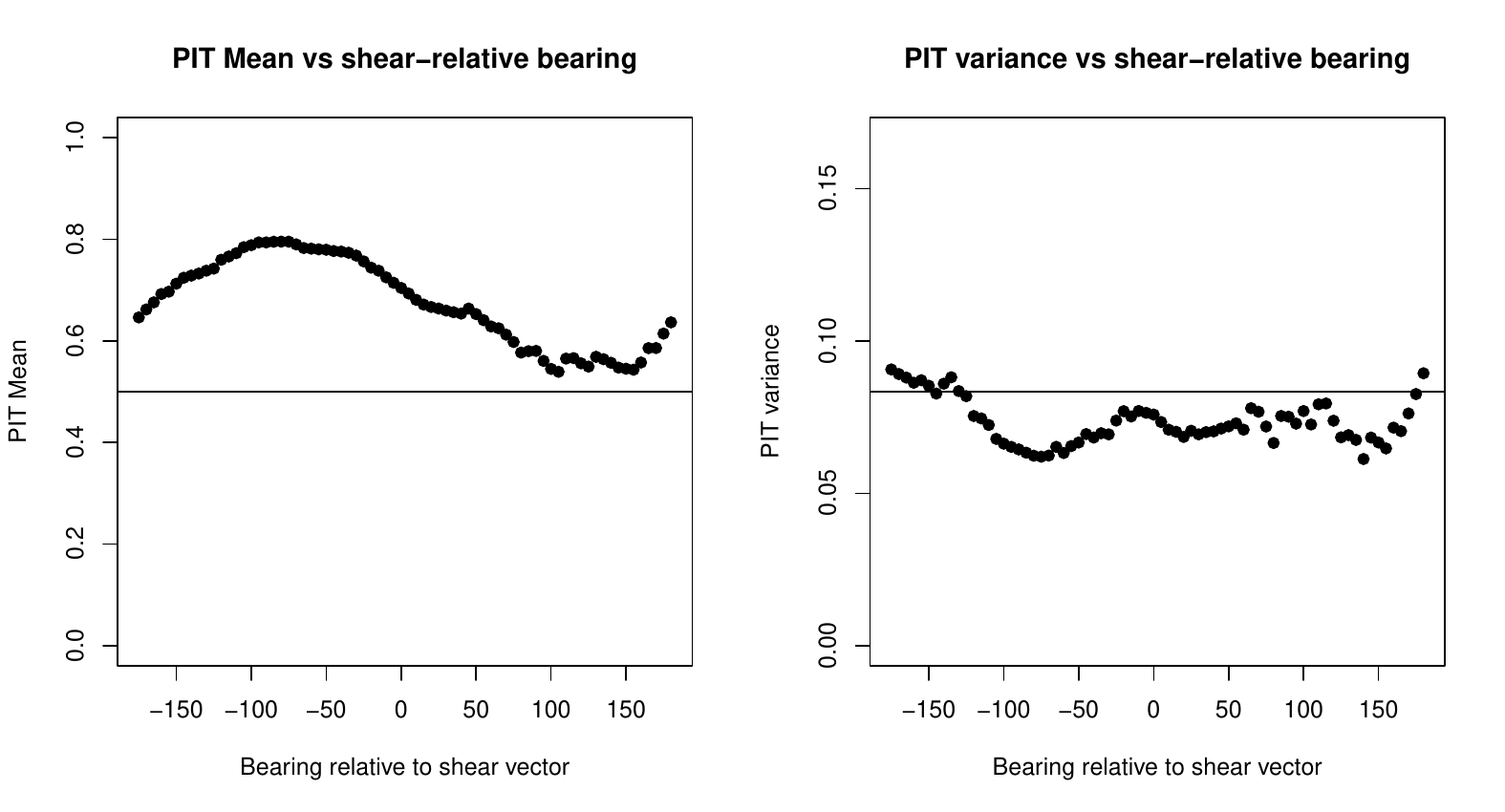}
    \caption{PIT mean and variance vs bearing relative to wind shear. Horizontal lines represent the mean and variance expected for a perfectly calibrated model. The results suggest that the model is systematically biased low, with the best calibration performance roughly right of the shear direction. }
    \label{fig:shear_plot}
\end{figure}

The model calibration was tested at the 6 different sites marked in Figure \ref{fig:storm_tracks_and_eval_locs} using 2500 realizations of the full model. The sites correspond to either cities along the Gulf Coast or oil fields in the Gulf itself. 
The resulting PIT histograms, shown in Figure \ref{fig:space_calibration}, show variable performance. While the calibration at Matagorda City, Boomvang Spar, and Corpus Christi is good, the model is clearly biased low in general, consistent with the low bias in the predictions of KED in Figure \ref{fig:ke_along_coast}. 
The variable calibration performance of this model is not necessarily surprising given its construction. One of the most important variables for tropical cyclone asymmetry and evolution, the wind shear, is left out of this model entirely \citep{TropicalCycloneWindShearRelativeAsymmetryinReanalyses,ObservedHurricaneWindSpeedAsymmetriesandRelationshipstoMotionandEnvironmentalShear}. While the random effects are meant to recover some of this 
unobserved storm-to-storm variation, they are assumed constant over the course of the storm and so cannot capture dynamically evolving effects such as those from a time-varying wind shear. To examine the relationship between the wind shear and model calibration, we calculated the wind shear from WRF as
\begin{equation}\label{eq:shear}
    \begin{split}
        u_{shear}&= u_{200mbar}-u_{850mbar}\\
        v_{shear}&= v_{200mbar}-v_{850mbar}\\
    \end{split}
\end{equation}
 averaged over the annulus of radii $100-400$ km for every time step of the model. We then calculated the PIT mean and variance for all spatial locations in shear-relative coordinates. The resulting average PIT mean and variance relative to the shear direction, as well as horizontal lines representing the values expected under perfect calibration, are plotted in Figure \ref{fig:shear_plot} and show very clear coherent structure as a function of direction relative to shear. Roughly, the model is best calibrated to the right of the shear direction as evidenced by the evolution of the PIT mean and worst calibrated in the direction left of shear, while the PIT variance is low except in the direction opposite the shear. However, regardless of direction, the model appears to be biased low relative to the WRF winds. These results suggest that exclusion of the wind shear from the model contributes to the miscalibration. The portion of this induced by shear cannot be easily accounted for unless those contributions are modeled explicitly. This behavior reflects a limitation of the separable construction of the mean model, which permits the magnitude of the radial wind profile to vary with azimuth but does not allow its radial structure to change with direction. Incorporating shear-dependent structure provides a natural extension of the model and is left for future work.

\section{Discussion}\label{sec:discuss_tcs}

One notable shortcoming of our approach is that the wind field is separable in $r$ and $\theta$, which is not realistic and likely contributes to issues with the calibration of the model. This is partially due to the limited nature of our dataset. However, more detailed track databases exist that contain information such as the radius of maximum winds in different directions. These data could be incorporated into this model, allowing characteristics of the radial profile to vary with azimuth at the cost of usability in more data limited cases. Wind shear provides another natural predictor of the asymmetry in the storm, and historical data are available from SHIPS \citep{demaria1999ships,shipsdevelopmentaldata}, although this again decreases the applicability of the model in data-limited settings. In the absence of this data, it may be possible to fit a separate model to the shear vector and take draws from it when simulating. 

Another alternative would be to abandon the parameterization described here entirely and instead describe the wind field via a Helmholtz decomposition as \citep{helmholtz_decomp}
\begin{equation}
    \label{eq:helmholtz_decomp}
    \begin{split}
        \vec{w}(\vec{x},t)&=\vec{w}_{bg}(\vec{x},t)+\vec{w}_{tc}(\vec{x},t)\\
        \vec{w}_{tc}(\vec{x},t) &= \vec{\nabla} \Phi(\vec{x},t) + \vec{\nabla}^{\perp}\Psi(\vec{x},t)
    \end{split}
\end{equation}
where $\Phi$ is the velocity potential and $\Psi$ the stream function, with the perturbation $\vec{w}_{tc}$ dropping to zero outside some radius $r$. These scalar functions solve the partial differential equations
\begin{equation}
    \label{eq:helmholtz_decomp_pdes}
    \nabla^2\Phi=\vec{\nabla} \cdot \vec{w}_{tc},\quad  \nabla^2\Psi=\vec{\nabla} \times \vec{w}_{tc}
\end{equation}
with appropriate boundary conditions. Depending on the numerical model, the potentials or the associated divergence and vorticity fields may be available directly, avoiding potentially noisy numerical differentiation of the wind fields necessary to solve the PDEs in Eq. \eqref{eq:helmholtz_decomp_pdes}. This naturally decomposes the field into the rotational (through $\Psi$) and irrotational (through $\Phi$) components. The problem then consists of discovering a statistical model of the two scalar fields, for example via a low-rank decomposition. A related approach was taken in \cite{tongyithesis} for modeling the lithospheric magnetic field using spherical needlets and in \cite{fan2018modeling} for generating random curl-free and divergence-free fields, and this formulation has already proven to be useful in the study of tropical cyclone formation in the Indian Ocean \citep{EMMANUEL2025185}.

Future research may consider the joint hazards due to TCs, for example by incorporating models of storm surge and precipitation along with the wind speed. 
The marginal behavior of precipitation has previously been modeled (see \cite{kleiber2023stochastic}), and the storm surge modeling may be done in several ways, for example by using this model to drive a storm surge model such as ADCIRC \citep{pringle2021adcirc}. Because the proposed model generates realistic ensembles of TC wind fields, it provides a natural framework for ensemble-based surge risk assessment and can be used with the model of \cite{kleiber2023stochastic} to develop a model for assessing the risk from extreme winds, storm surge, and precipitation. 




\section{Disclosure statement}\label{disclosure-statement}

The authors have no conflicts of interest to report.

\section{Acknowledgements}

This paper describes objective technical results and analysis. Any subjective views or opinions that might be expressed in the paper do not necessarily represent the views of the U.S. Department of Energy or the United States Government. Sandia National Laboratories is a multimission laboratory managed and operated by National Technology \& Engineering Solutions of Sandia, LLC, a wholly owned subsidiary of Honeywell International Inc., for the U.S. Department of Energy’s National Nuclear Security Administration under contract DE-NA0003525.

\newpage
\bibliographystyle{imsart-nameyear} 
\bibliography{refs_sandia.bib,refs_needlets.bib,refs_spatialcov.bib}

\end{document}